\documentclass[fleqn,usenatbib]{mnras}

\usepackage{newtxtext,newtxmath}

\usepackage{longtable}

\usepackage[T1]{fontenc}

\usepackage{lscape} 
\usepackage{array}

\DeclareRobustCommand{\VAN}[3]{#2}
\let\VANthebibliography\thebibliography
\def\thebibliography{\DeclareRobustCommand{\VAN}[3]{##3}\VANthebibliography}

\usepackage{graphicx}	
\usepackage{amsmath}	

\title[Properties of protoplanetary discs]{Fundamental properties of protoplanetary discs determined from simultaneous fits to thermal dust images and spectral energy distributions} 

\author[Tim J. Harries]{
Tim J. Harries\thanks{E-mail: t.j.harries@exeter.ac.uk (TJH)}
\\
Department of Physics \& Astronomy, University of Exeter, Stocker Road, Exeter EX4 4QL, UK\\
}

\date{Accepted 2026 March 15. Received 2026 March 5; in original form 2026 January 9}

\pubyear{\the\year{}}

\begin{document}
\label{firstpage}
\pagerange{\pageref{firstpage}--\pageref{lastpage}}
\maketitle

\begin{abstract}
We present a novel machine learning method that is capable of rapidly and accurately producing dust-continuum model images and spectral energy distributions from training sets created using a detailed radiative transfer code. We create a training set that encompasses the parameter space for protoplanetary discs, and then couple the trained machine learning method with a Bayesian optimisation algorithm. We then simultaneously fitted 1.3\,mm ALMA ODISEA survey images of protoplanetary discs in the Ophiuchus Molecular Cloud, and their spectral energy distributions, in order to determine fundamental discs parameters such as dust masses and radii. We find that good simultaneous fits may be found for the Class~II objects in the survey,  although the spectral fits are poorer for the Class~I and flat spectrum sources. We find that the dust mass distributions of discs is broader and shallower than that predicted from 1.3\,mm flux dust mass estimates, substantially increasing the numbers of objects with high-mass and low-mass discs. We show that this is due to a combination of optical depth and dust temperature effects, which are strongly related to the disc size and inclination constraints provided by the imaging fits. We show that there is a significant decrease in disc scale height and disc flaring when moving from the the Class I objects, to the flat spectrum sources, and the Class II discs.
\end{abstract}




\section{Introduction}

Protoplanetary discs are the birthplace of new planetary systems, and understanding the conditions of dust and gas in these discs is key to making sense of the configurations of exoplanetary systems \citep{Miotello2023}. One of the most basic quantities we wish to know is how much dust each protoplanetary disc contains, but unfortunately determining this parameter with certainty is proving challenging. 
As a first approximation it is often assumed that the emitting dust is optically thin and at a constant temperature ($T_{\rm dust}$). This enables the use of the sub-millimetre flux $F_\lambda$ as a proxy for the dust mass:
\begin{equation}
M_{\rm dust} = \frac{F_\lambda d^2}{\kappa_\lambda B_\lambda(T_{\rm dust})}
\label{eq:mdust_flux}
\end{equation}
where $d$ is the distance, $\kappa_\lambda$ is the dust opacity, and $B_\lambda$ is the Planck function \citep{Beckwith1990}. This elegant formulation enables observers to simply estimate the dust mass, and it should be insensitive to the precise geometrical distribution of the dust. Broadly speaking this approach has remained unchanged for several decades, albeit with some small modifications, such as scaling $T_{\rm dust}$ by the effective temperature of the protostar \citep{Andrews2013}. The advance in the observational capacity to observe in the sub-mm provided by ALMA has led to a host of sub-mm surveys of dust in nearby star forming regions, enabling observers to examine the evolution and environmental dependence of the dust mass of protoplanetary discs (see review by \citealt{Manara2023}). 

Despite these successes, there is growing evidence that the main assumption the underpins sub-mm dust mass estimates (that the dust is optically thin to sub-mm radiation) isn’t necessarily true and that self-scattering by the emitting dust may be significant. This means that dust masses will be underestimated, perhaps by an order of magnitude in certain circumstances (e.g. \citealt{Zhu2019}; \citealt{Ballering2019}; \citealt{Ueda2020};  \citealt{Liu2022}; \citealt{Xin2023}). A more sophisticated approach is therefore necessary in order to accurately model the radiative transfer and thermodynamics of the disc.

Despite the widespread use of sub-mm flux mass estimates, modern radiative transfer (RT) codes, predominantly based on Monte Carlo (MC) methods, are routinely used to model protoplanetary discs. These simulations self-consistently account for the self-scattering and should produce a much more reliable dust estimate. Until recently this required fitting the spectral energy distribution (SED), usually via a small grid of bespoke models created on an object-by-object basis e.g. \cite{Pinte2008}. This is a time-consuming business, as a typical RT model might take several minutes to run on a multi-core computer, and many thousands of models are needed to fit an SED and explore parameter degeneracies.  However deep learning methods, using artificial neural networks (NN), may be used to speed up this process: Given a set of training data, and appropriately sampling disc parameter space, it is possible to take the disc parameters as inputs to a NN and the monochromatic fluxes of the model SED as outputs. The NN learns the complex non-linear, multi-dimensional relationship between the inputs and the outputs, and subsequently can produce an SED from an arbitrary set of disc parameters many orders of magnitude more quickly than the RT code.  By coupling the trained NN with an optimiser it is possible to fit extant survey data of discs in nearby star-forming regions and use Markov chain Monte Carlo methods to explore parameter degeneracies. Studies that use this technique are starting to appear in the literature (e.g. \citealt{Ribas2020}; \citealt{Kaufer2023}; \citealt{Rilinger2023}).

The ability to fit SED surveys is a powerful development, but disc parameters are not uniquely determined by fitting the SED alone, and a much tighter grip on parameter space may be exerted by employing other observables. ALMA surveys for example often provide size constraints on the disc thermal emission which helps constrain the disc radius. \cite{Ballering2019} demonstrated that by incorporating sub-mm disc-size constraints in tandem with SED fitting much more reliable dust mass estimates can be obtained. 

In this paper we extend the use of machine learning approaches to enable the  generation of both SEDs and sub-mm imaging using a NN trained on models produced using a state-of-the-art radiative transfer code. By coupling the NNs with objective fitting methods were are able to simultaneously fit SEDs and imaging, thus producing reliable dust estimates for large surveys based on detailed RT modelling. 

In the following we first describe the neural network configurations that we use, and the production of our training data set. We subsequently apply the new technique to the ODISEA survey of protoplanetary discs in $\rho$~Oph. We compare the resulting dust masses with those derived from 1.3\,mm fluxes, and examine the role of optical depths on sub-mm dust mass estimates. Finally, we give our perspectives on the future applications of the method in enhancing not only our understanding of dust masses on a statistical basis but also our detailed knowledge of individual objects by simultaneous fitting of spectral, imaging and visibility data, from the dust continuum as well as spectral lines.

\section{The machine learning algorithms}

Artificial neural networks rely on layers of neurons. In the simplest case of a fully connected, feed forward network an individual neuron in receives inputs from all neurons in the previous layer, and passes its output to all neurons in the subsequent layer. Consider an individual neuron $i$ in layer $\ell$. It is associated with a bias value $b_i$ and a set of weights $w_{i,j}$ which are used to weight the outputs $a_j$ of the neurons from the previous layer ($\ell-1$) which contains $N_{\ell-1}$ neurons. Thus the input to our individual neuron is given by
\begin{equation}
x_i = \sum^{N_{\ell-1}}_{j=1} w_{i,j} a_{\ell-1,j} + b_i
\end{equation}
If $x_i$ is taken to be the output of the neuron i.e. $a_{\ell, i} = x_i$ then it has linear dependence on the inputs, and thus the final network output becomes a linear combination of linear functions and is thus itself linear. This would not be a particularly useful NN. We must introduce some non-linearity into the system, and this is done using an activation function, $a(x)$. There is a host of activation functions in widespread use, but in this paper we use two different activation functions, the first of which is the sigmoid function
\begin{equation}
a(x) = \sigma (x) = \frac{1}{1+e^{-x}}
\end{equation}
which tends to zero as $x \rightarrow -\infty$ and unity as $x \rightarrow +\infty$. The second activation function we use is the rectified linear unit (abbreviated to ReLU), a positive ramp:
\begin{equation}
a(x) = {\rm ReLU}(x) = {\rm max}(0,x) = \frac{x + |x|}{2}.
\end{equation}
The output for our specific neuron $i$ is therefore $a(x_i)$, and this becomes one of the inputs to neurons in layer $\ell+1$. The goal is to use a training dataset to determine the set of weights $w_{i,j}$ and biases $b_i$ which best reproduces the training dataset given the training inputs. The goodness-of-fit quantity that measures the level agreement between the output of the NN given the input parameters is termed the loss function, and can be a traditional quantity such as the mean-square error or the mean absolute  deviation. The training of the neural network is performed using the back propagation algorithm \citep{Rumelhart1986}, which relies on an estimate of the local gradient of the loss function with respect to the biases and the weights. Rather than using the entire training dataset to calculate this gradient it is approximated using a subset, or batch, of the training set. For each batch the gradient is calculated and the weights and biases adjusted in the direction of the negative gradient, and thus towards the minimum of the loss function. This optimisation, the simplest of many optimisation algorithms, is known as stochastic gradient descent (SGD). An epoch of training is complete when every batch comprising the training data has been used. Many epochs of training are required for the network to be considered sufficiently trained. 

The design of the network is an important step, and its configuration and properties are described as the hyperparmeters of the network which include its design (the number of nodes, the number of layers, the activation functions) as well as its ancillary properties used in training such as the batch size and the optimiser. In order to facilitate the network design the dataset is usually split into two components. The largest is the training dataset which is used to optimised the weights and biases. The second is the validation set, which is used to assess the efficacy of the training and aid optimisation of the network configuration.

\subsection{SEDs}
\begin{figure}
	\includegraphics[width=\columnwidth]{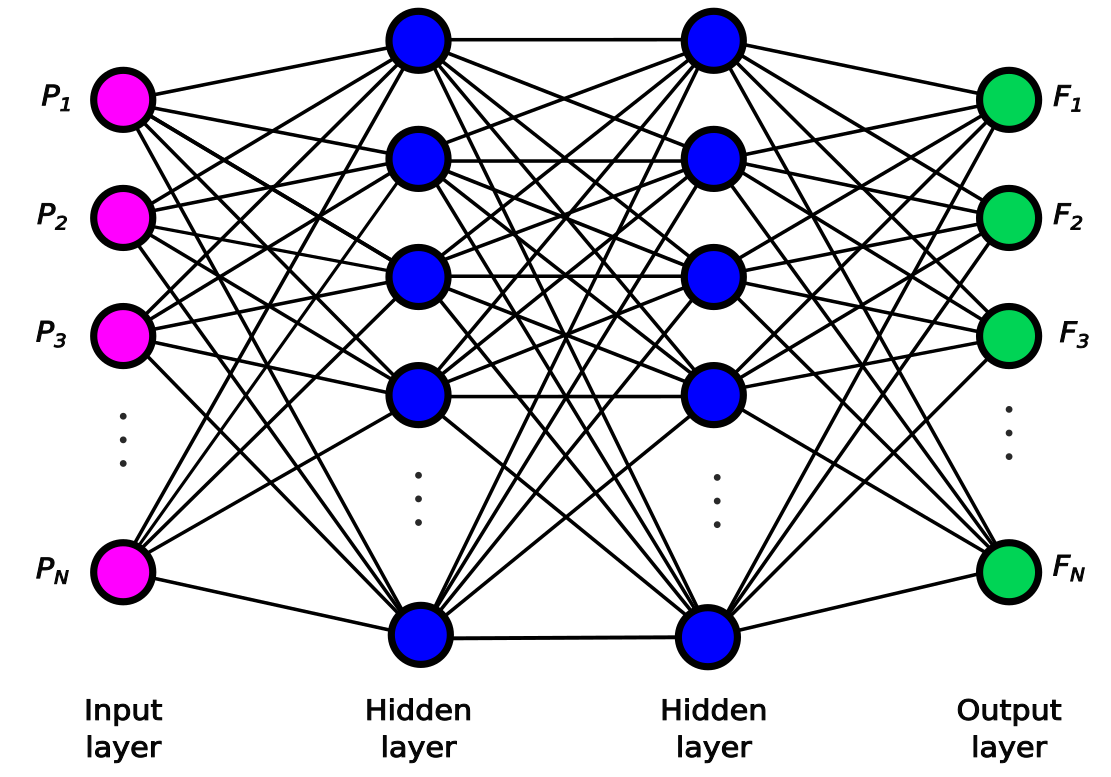}
    \caption{A schematic of the feed-forward, fully-connected neural network used to predict the model SEDs. The input layer (pink neurons) consists of the normalised model inputs $P_i$. The two dense (fully connected) hidden layers are shown in blue, with the final SED flux outputs ($F_i$) in green.}
    \label{fig:sed_nn}
\end{figure}

Reproducing the SEDs requires computing $\sim 100$ flux values from $\sim 10$ disc parameters, and a feed-forward, fully-connected NN is ideally suited to such a problem. For example, \cite{Ribas2020} adopted a NN with two hidden layers of 250 neurons, combined with rectified linear unit (ReLU) activation functions. \cite{Kaufer2023} successfully employed a similar (but much deeper) network with six hidden layers of 300 neurons each (comprising nearly 500\,000 free parameters) to model disc SEDs. 

We also employ fully connected, feed forward network. We implemented the network using the {\sc keras} python module \citep{chollet2015keras}, which is part of the {\sc tensorflow} library \citep{tensorflow2015-whitepaper}. Following \cite{Ribas2020} we adopted two hidden layers between our disc parameters and our 100 SED fluxes, with ReLU activation functions. A schematic of the network is given in Figure~\ref{fig:sed_nn}. We adopted the mean absolute error as the loss function, and used the Adam optimiser to train the network. We provide details on how we optimised the network hyperparameters in Appendix~\ref{appendix_a}.

\subsection{Images}

One might naively think that a fully connected feed-forward neural network, which works so well for the SED problem, might be equally serviceable for the problem of generating images. Indeed \cite{Williams2025} used such a network, with two hidden layers of 40 and 400 neurons and leaky ReLU activation functions, to calculate scattered light images. However \cite{Telkamp2022} had difficulty reproducing scattered light images of edge-on disks with sufficient accuracy using this method, this is likely due to the larger size of the images considered ($187 \times 187$ pixels versus $40 \times 40$ in the Williams et al. case). The fundamental drawback of using a fully-connected neural network to predict images is that it neglects the correlation between adjacent pixels that naturally occur in images, and in the general case of large images vast numbers of neurons (and therefore free parameters) must be used, leading to problems with training and convergence. A better approach is to use a network that employs 2D convolutions, which can reduce the number of free parameters of a model and allow networks to learn the repeating patterns that are inherent in imaging data. Such networks are based on convolutions, and were used with some success to model edge-on discs by \cite{Telkamp2022}.

A 2D convolutional layer consists of a set of learnable patch filters that are convolved with the image according to a stride pattern, to produce a corresponding set of feature maps that encode regular features of the images (such as corners and edges, or other regularly occuring patterns). Several convolutional layers may be applied in series to reduce the dimensions of an image. The inverse of the convolutional layer is the convolutional transpose, which employs filters to expand the image dimensions. 

A promising network configuration for the problem at hand is the autoencoder (AE), in which convolutional layers are used to reduce the image dimensions to a flattened vector termed the latent space, which then feeds into a set of transpose layers to recover an approximation of the original image. This bottleneck configuration was first proposed by \cite{Kramer1991}. Training the autoencoder on a set of images yields a network which can be useful for image compression and denoising, but more importantly for this research, once the autoencoder is trained it is possible to generate novel images directly from the latent space, bypassing the encoder entirely. 

For the specific case of protoplanetary disc images we expect the latent space to encode information about the images in the usual way,  and crucially that the latent space will be intimately related to the input parameters of the disc model, such as the disc mass and inclination. This in itself might be considered a naive approach, since it is well known that the latent space of an autoencoder can be disordered and rather arbitrary, with discontinuities and voids that can render simple interpolation challenging. Nonetheless our hope is that if we train an autoencoder on a sufficiently large number of images we can create a large set of latent space vectors whose distribution does encode a relatively ordered representation of the images. We can then train another network (in this case a feedforward fully-connected network) to learn the relationship between the disc parameters and the latent space. Combining this network with the decoder section of the autoencoder we can then produce disc images for arbitrary sets of disc parameters. 

There is a wide range of possible network configurations for the autoencoder, including the size of the patch filters, the number of filters, the number of convolutional layers, the size of the latent space, and so on. We provide details on the range of hyperparameters we examined in the Appendix~\ref{appendix_a}, and give our final choices for these parameters here.

\begin{figure*}
	\includegraphics[width=180mm]{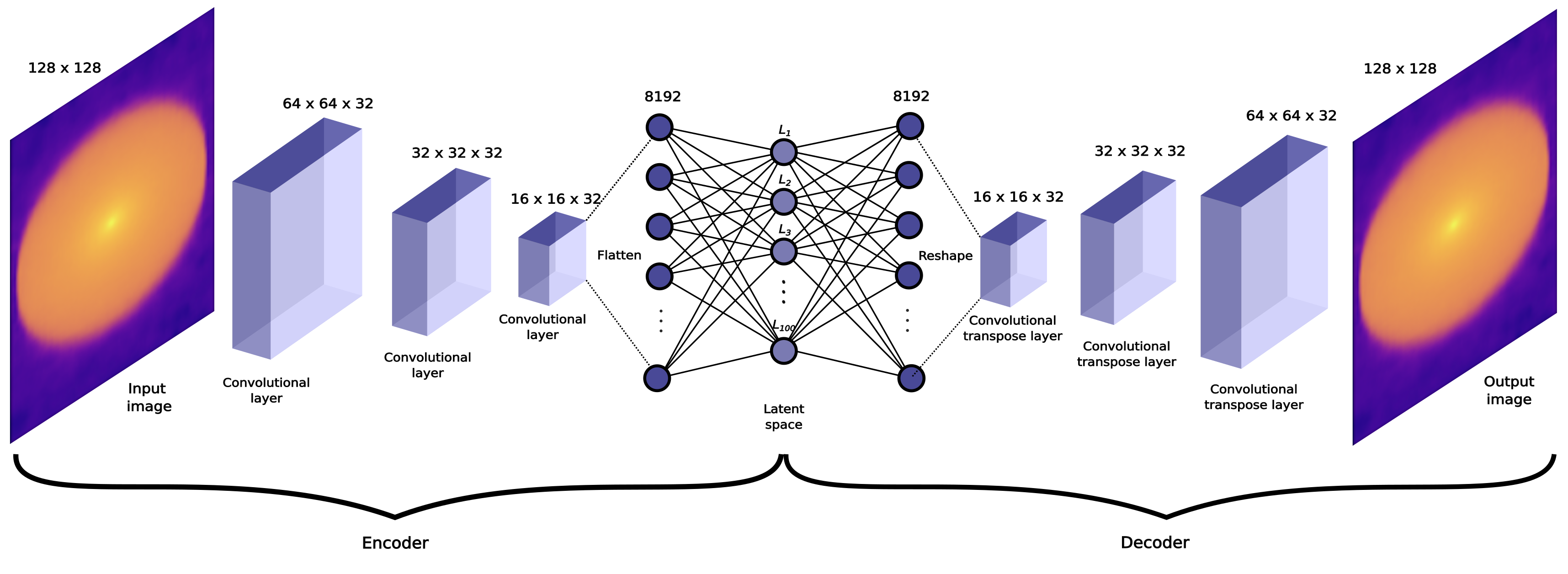}
    \caption{A schematic of the autoencoder configuration. The original image (left hand panel) passes through three convolutional layers to reduce its size from $128 \times 128$ pixels to $16 \times 16$. The final convolutional layer is flattened and fully connected to a 100 neuron latent space ($L_1$ to $L_{100}$). Subsequently the reverse process is applied using convolutional transpose layers to recreate an approximation of the original image.}
    \label{fig:autoencoder}
\end{figure*}

The encoder part of the AE takes the image as the input layer, and uses three convolutional layers to reduce the dimension of the image from $128 \times 128$ pixels to $16 \times 16$ pixels. Each layer consists of 32 $3 \times 3$ filters that are applied with a stride of two. The ReLU activation function was adopted in each layer. The $16 
 \times 16$ layer is flattened to one dimension and is then fully connected to a 100 neuron latent space layer, which is the final output of the encoder part of the network, and the bottleneck of the autoencoder. For the decoder the 100 neuron latent space is connected to a $16 \times 16 \times 32$ neuron layer which is then reshaped to 32 $16 \times 16$ images (feature maps). Subsequently three 2-d convolutional transpose layers expand the image back to the $128 \times 128$ original size. A final layer applies a sigmoid activation function (appropriate as the final pixel values will have values between 0 and 1). A schematic of the autoencoder configuration, demonstrating its bottleneck structure, is given in Figure~\ref{fig:autoencoder}. In order to train the autoencoder we used the mean squared deviation as the loss function, and minimized this using the Adam optimiser and an initial learning rate of 0.0001. After 100 epochs we found that the network had converged sufficiently. 

\begin{figure*}
	\includegraphics[width=180mm]{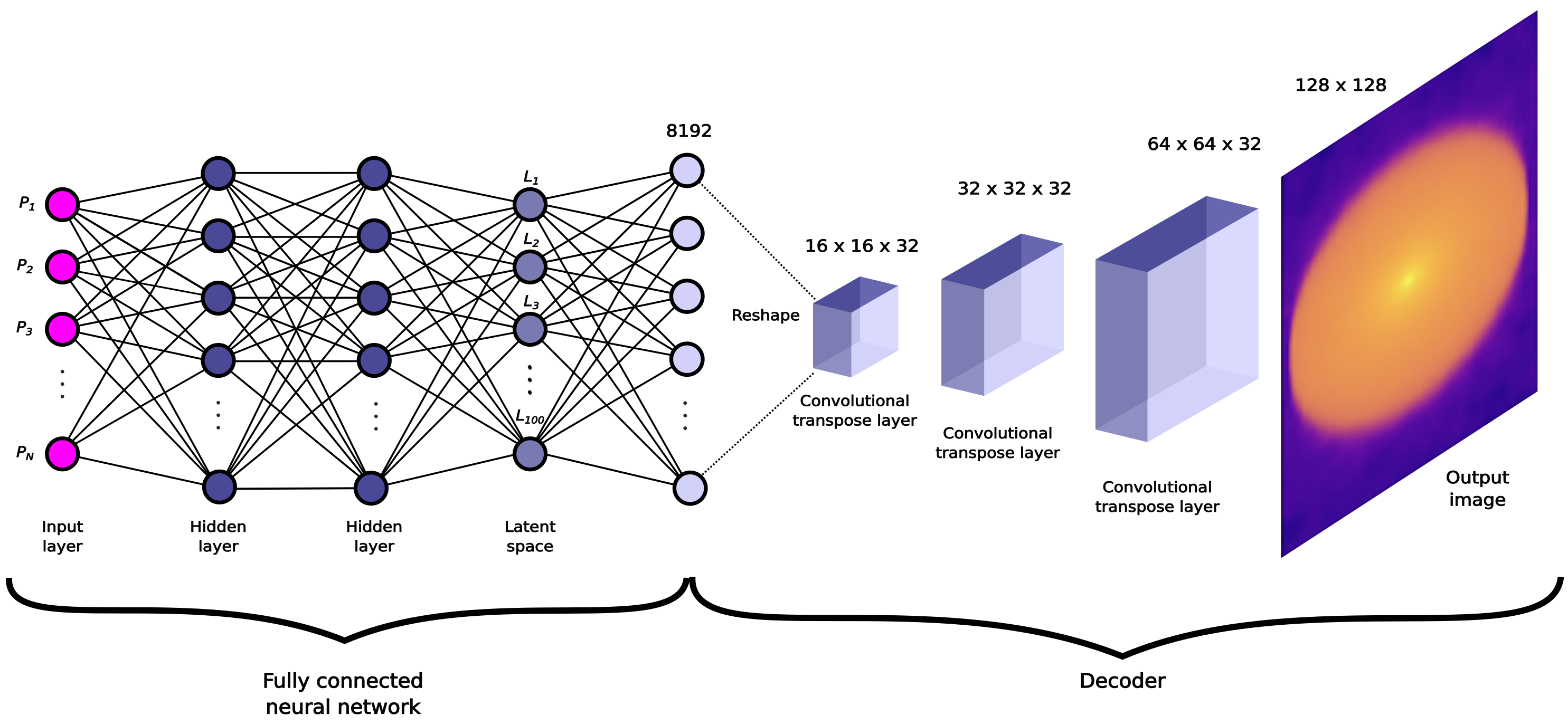}
    \caption{A schematic of the decoder network. The inputs (pink symbols) are the normalised disc parameters $P_i$ (such as mass, inclination, radius etc), which are fully connected to two hidden layers (dark grey symbols) and then to the latent space ($L_i$, light grey symbols). The right-hand part of the schematic represents the decoder part of the (previously trained) autoencoder.}
    \label{fig:decoder}
\end{figure*}

Once the autoencoder is trained we have a large dataset of disc parameters and corresponding latent space vectors. We use fully-connected NN to learn the relationship between the disc parameters and the latent space representations, using a two-layer network of 500 neurons each, connected by ReLU activation functions and adopting the mean squared error as the loss function. (Once again our investigation of the choice of hyperparameters is given in Appendix~\ref{appendix_a}). The output of this network (which is a latent space vector) is then passed to the decoder part of the autoencoder, which results in an output model image.
A schematic of the complete latent space/decoder network is given in Figure~\ref{fig:decoder}.

\section{Training data}

The training data (images and SEDs) were calculated using the {\sc torus} Monte Carlo radiative transfer code. The code uses the Monte Carlo method and the radiative equilibrium algorithm of \cite{Lucy1999}, and is described in detail in \cite{Harries2019}.

\subsection{The disc model}

For the purposes of this study we assume that the disc is a continuous structure that extends from an inner (cylindrical) radius $r_{\rm in}$ to and outer radius $r_{\rm out}$. The scale height of the gas is assumed to follow a power law
\begin{equation}
h(r) = h(r_{\rm out}) \left( \frac{r}{r_{\rm out}} \right)^\beta
\label{eq:gas_height}
\end{equation}
and the gas density declines as a power law in $r$ and an exponential in cylindrical height $z$ according to
\begin{equation}
\rho(r,z) = \rho_0 \left( \frac{r}{r_{\rm in}} \right)^{-\alpha} \exp \left( -\frac{1}{2} \left( \frac{z}{h(r)} \right)^2 \right)
\end{equation}
The value of $\rho_0$ is fixed by and integration of the density over volume to get the disc mass
\begin{equation}
M_{\rm disc} = \int_{r_{\rm in}}^{r_{\rm out}} \int_{-\infty}^{\infty} 2 \pi r \rho(r,z)\,dz\,dr.
\end{equation}
For convenience we combine $\alpha$ and $\beta$ to give a surface density power law index 
\begin{equation}
\Sigma(r)  = \Sigma_0 \left( \frac{r}{r_{\rm in}} \right)^{-\gamma}
\end{equation}
where $\gamma = \beta - \alpha$ and the fiducial surface density $\Sigma_0$ is fixed in the same manner as $\rho_0$.

\subsection{The dust model}

We assume that the dust grains have a size distribution 
\begin{equation}
n(a) \propto a^{-q}
\end{equation}
where $n$ is the number of grains of size $a$ and $q$ is a power law index (which is fixed to $3.5$ in this work). We adopt a minimum grain size of 5\,nm and a maximum grain size of 1\,mm. The dust is assumed to be 67\% silicates and 33\% carbonaceous grains by mass, with the optical properties taken from \cite{Draine1984} and \cite{Zubko1996}. We allow grain settling according to the prescription of \cite{Pinte2008}, in which the scale height of the dust density is given
\begin{equation}
h(a) = h_0(a_{\rm min}) \left( \frac{a}{a_{\rm min}} \right)^{-\xi}
\end{equation}
where $h_0(a_{\rm min})$ is equal to the gas scale-height (equation~\ref{eq:gas_height}). We assume a dust-to-gas ratio of 1:100 by mass.

Note that the 1.3\,mm dust opacity from this distribution is 5.3\,cm$^2$\,g$^{-1}$ and we calculate the 1.3\,mm dust mass estimates using this opacity, rather than the 2.25\,cc$^{2}$\,g$^{-1}$ employed by \cite{Williams2019} in their study of dust masses from the ODISEA survey. This allows for a like-for-like comparison between the dust mass estimates from the 1.3\,mm flux dust mass estimates and those from the {\sc torus} models.

\subsection{The stellar model}

We opted to calculate the stellar parameters by sampling in stellar mass and age and adopting the MIST pre-main sequence evolutionary models \citep{Dotter2016,Choi2016} to obtain the corresponding stellar radii and effective temperatures. We used the python module {\sc pystellibs} by M.\,Fouesneau\footnote{\tt https://mfouesneau.github.io/pystellibs/} to interpolate in solar-metallicity model atmosphere grids to obtain the appropriate stellar spectrum. For a particular effective temperature ($T_{\rm eff}$) and surface gravity ($\log g$) we used the TLUSTY models \citep{Lanz2003} if available and required for our hottest stars, or alternatively and more usually the BT-Settl Library \citep{Allard2000}. We adopted the ATLAS9 models \citep{Castelli2003} if the $T_{\rm eff}$, $\log g$ combination fell outside both the TLUSTY and BT-Settl ranges.

\subsection{The training grid}

The range of disc model parameters used are listed in Table~\ref{tab:parameters}, along with the type of sampling (either linear or logarithmic) adopted. The model parameters were sampled using Sobol pseudo-random numbers \citep{Sobol1967}, which provides a more even sampling of the parameter ranges than strict pseudo-random sampling. We note that the minimum value of the inner radius is defined by the dust sublimation radius. Although {\sc torus} is capable of self-consistently calculating the dust sublimation front, this is a significant computational overhead, so for the purposes of the training grid we used the approximation
\begin{equation*}
r_{\rm sub} = 0.06 \left( \frac{T_{\rm eff}}{4000} \right)^2 \left( \frac{R_*}{2} \right)
\end{equation*}
where $r_{\rm sub}$ is the sublimation radius in au, and $R_*$ is expressed in solar units \citep{Woitke2015}.
\begin{table}
	\centering
	\caption{Model parameters used to construct the training data.}
	\label{tab:training_grid_parameters}
	\begin{tabular}{lccr} 
	\hline
	Model parameter & Units & Range/Value & Sampling\\
	\hline
  
    \multicolumn{4}{c}{Stellar properties} \\
    
	Stellar age & years & $10^5$--$10^7$ & log \\
    Stellar mass, $M_*$  & $M_\odot$ & $0.5$--$4$ & linear \\
    \hspace*{2mm} \\
    \multicolumn{4}{c}{Disc properties} \\
	Dust mass, $M_{\rm dust}$ & $M_\odot$ & $10^{-6}$--$10^{-4}$& log 
\\
    Inner radius, $r_{\rm in}$ & au & $r_{\rm sub}$ -- 0.5 & linear \\
    Outer radius, $r_{\rm out}$ & au & 10 -- 300 & linear \\
    $h/r$ (defined at $r_{\rm out}$) &  & 0.05 -- 0.2 & linear \\
    Height power-law index, $\beta$ & & 1 -- 1.5 & linear \\
    Radial suface density index, $\gamma$& & 0.5 -- 2 & linear \\
    Inclination, $i$ & deg & 0 -- 180 & $\cos i$
    \hspace*{2mm} \\
    \multicolumn{4}{c}{Dust properties} \\
    Maximum grain size, $a_{\rm max}$ & mm & 1  & fixed\\
    Minimum grain size, $a_{\rm min}$ & nm & 5 & fixed \\
    Grain size power law, $q$ & & $-3.5$ & fixed \\
    Mass fraction of silicates &  & 0.665 & fixed \\
    Mass fraction of amorphous carbon & & 0.335 & fixed \\
    Dust settling parameter, $\xi$ & &  0 -- 0.2 & linear \\
		\hline
	\end{tabular}
\label{tab:parameters}
\end{table}For each set of parameters the code first sets up a 2D-cylindrical computational domain, using an adaptive mesh to sample the opacity gradients appropriately. The radiative equilibrium method is then applied iteratively until convergence (which typically occurs after 5 iterations). The dust temperatures are then used to calculate the emissivities and hence the spectral energy distributions (at 100 logarithmically spaced wavelengths between 0.1 and 1300\,$\mu$m) and the dust continuum images ($128 \times 128$ pixels, with a linear size of 2.2\,$r_{\rm out}$. The SEDs and images of a particular model are computed at three random inclinations, which is a compromise between producing as many independent parameter samples as possible whilst making efficient use of the computational expense of the radiative equilibrium step.

We ran the models on University of Exeter's HPC facility, yielding a total of 39297 SEDs and images (13099 disc models viewed at three inclinations) which were were divided into 80:20 into training and validation datasets. Note that it is tempting to view this method as a multi-dimensional interpolation. This is not a strictly accurate interpretation, firstly since of course an interpolating function is designed to pass through all the nodes (which at least in this case the NN does not guarantee) but primarily because we can see that the number of models we have computed is considerably smaller than would be required for, say, evenly-spaced 5 point sampling of the parameter ranges over our 10 free parameters (which would be around 10\,million models). Nonetheless even this relatively modest training set is perfectly adequate for these neural networks, which can produce excellent SEDs and images for arbitrary disc parameters as we see describe in the following section.

\section{Training the networks}

It is important to normalise the input and output data for the neural networks since this aids the numerical calculations and the convergence of the network. The inputs, which consists of the disc model parameters, were normalised by subtracting the mean and dividing by the standard deviation of the parameters calculated over all training models.  We examine the training of the SED and image networks separately.

\subsection{The SED network}

The network is trained on logarithmic flux, normalized by subtracting the mean and dividing by the standard deviation calculated for each monochromatic flux over the training dataset. The training was performed using the Adam optimiser \citep{Kingma2015} with a batch size of 64 and an initial learning rate of 0.0001. The training was run for 400 epochs.

\begin{figure}
	\includegraphics[width=\columnwidth]{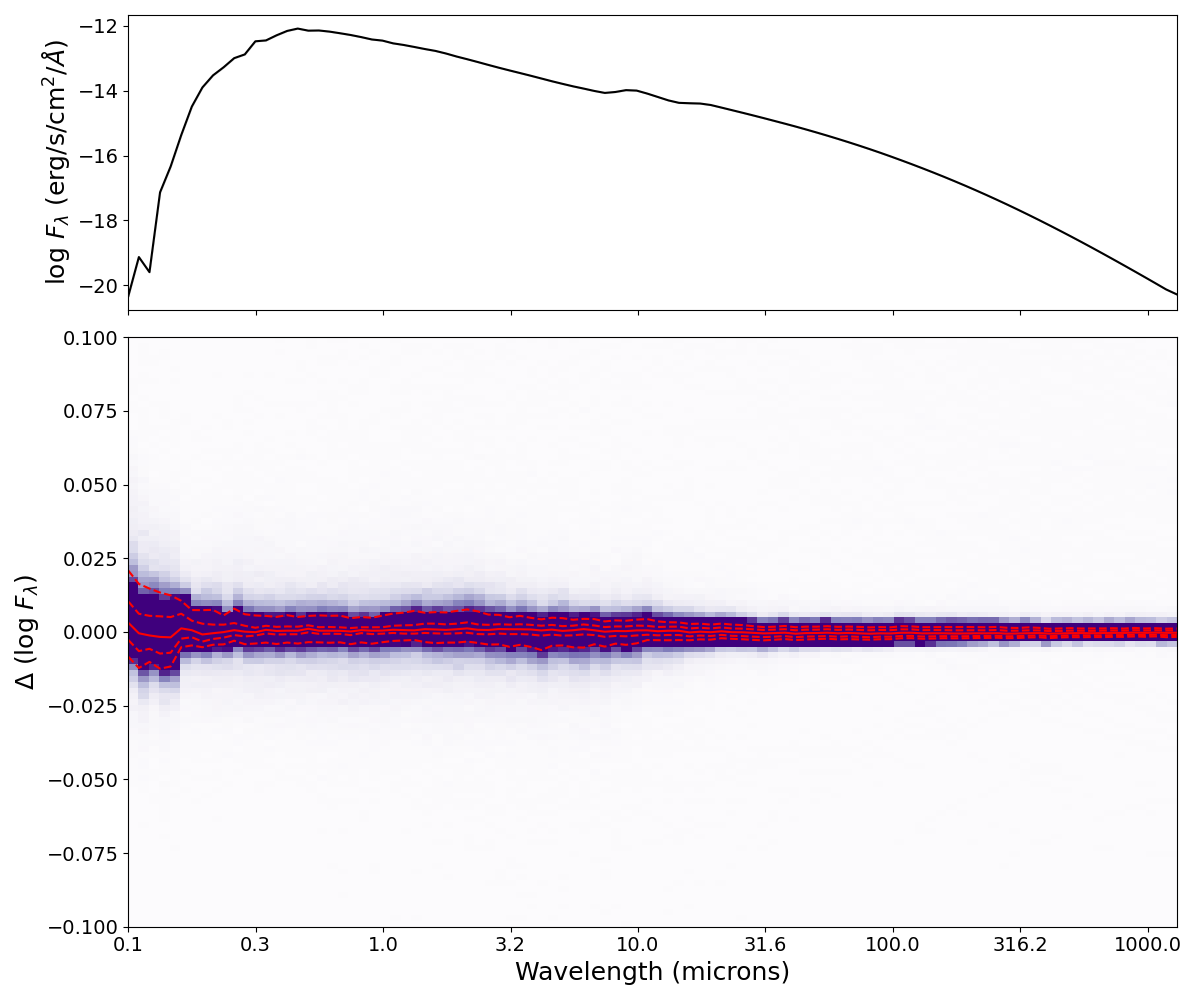}
    \caption{The top panel shows the mean SED of the test dataset. The bottom panel shows the histogram of the mean absolute differences between the validation data and the output of the neutral network for each wavelength bin (colour scale). The mean (solid read line), 1-sigma and 2-sigma limits (dotted lines). }
    \label{fig:hist_2d_fig}
\end{figure}

We illustrate the accuracy of the SED generating NN in Figure~\ref{fig:hist_2d_fig}, which shows the mean SED along with a 2D histogram of the  differences in the log between the validation data and the neural network for each wavelength bin.  Our final network reproduces the validation data to a standard deviation in log space of $\pm 0.01$ (2-sigma) across the entire spectral range. The standard deviation is smallest at long wavelengths, and grows slightly at short wavelengths (where the noise in the training data due to the Monte Carlo nature of the calculations is largest, particularly for edge-on discs). These errors are well below typical random and systematic uncertainties on the photometry of protoplanetary discs, giving us confidence that we can use the network-derived SEDs for objective fitting of observations.

\subsection{The imaging network}

The images in surface brightness have a large dynamic range, both within the image itself but also across the images within the training set, where the dust mass alone varies by orders of magnitude. We therefore chose to normalise all the images to their maximum values. The disadvantage of this normalisation is the loss of the flux constraint from the imaging, but this is mitigated by the fact that we are simultaneously fitting both the images and the spectral data (which contains the 1.3\,mm flux point). The considerable dynamic range within each image pointed us towards using logarithmic brightness within the image, so we adopted the following normalization for the scaled pixel brightness, $p_{i,j}$:
\begin{equation}
p_{i,j} = \frac{1}{5} \left[ 5 + {\rm max} \left( -5,\log_{10} \left( \frac{f_{i,j}}{{\rm max}(f_{i,j})} \right) \right) \right]
\end{equation}
where $f_{i,j}$ is the model surface brightness. This formulation means that $p_{i,j}$ is normalised and logarithmic brightness between the maximum brightness and five orders of magnitude fainter than the brightest pixel (this is conservative cut-off, and is much deeper than the noise level of the ALMA data). We found that this scaling gave better results than a simple linear normalization. 

We trained the AE using the Adam optimiser and an initial learning rate of 0.0001, with a batch size of 32. We found that the network converged after 100 epochs. Following this we trained the NN on the disc parameters to learn the latent space distributions, using a similar training approach. (Note that once again we applied a normalisation to the latent space itself). 

The question is how well does the AE network (and the latent space representation of the images) reproduce the validation set? Furthermore how well do the images generated using the decoder network reproduce the validation images? It is possible that the AE works well but that the latent space is not sufficiently constrained to make a reliable connection between it and the original disc parameters. We have trained the networks on the pixel-by-pixel mean squared deviation as our loss function, but it is possible that this is not a robust measure of the quality of the images.

\begin{figure}
	\includegraphics[width=\columnwidth]{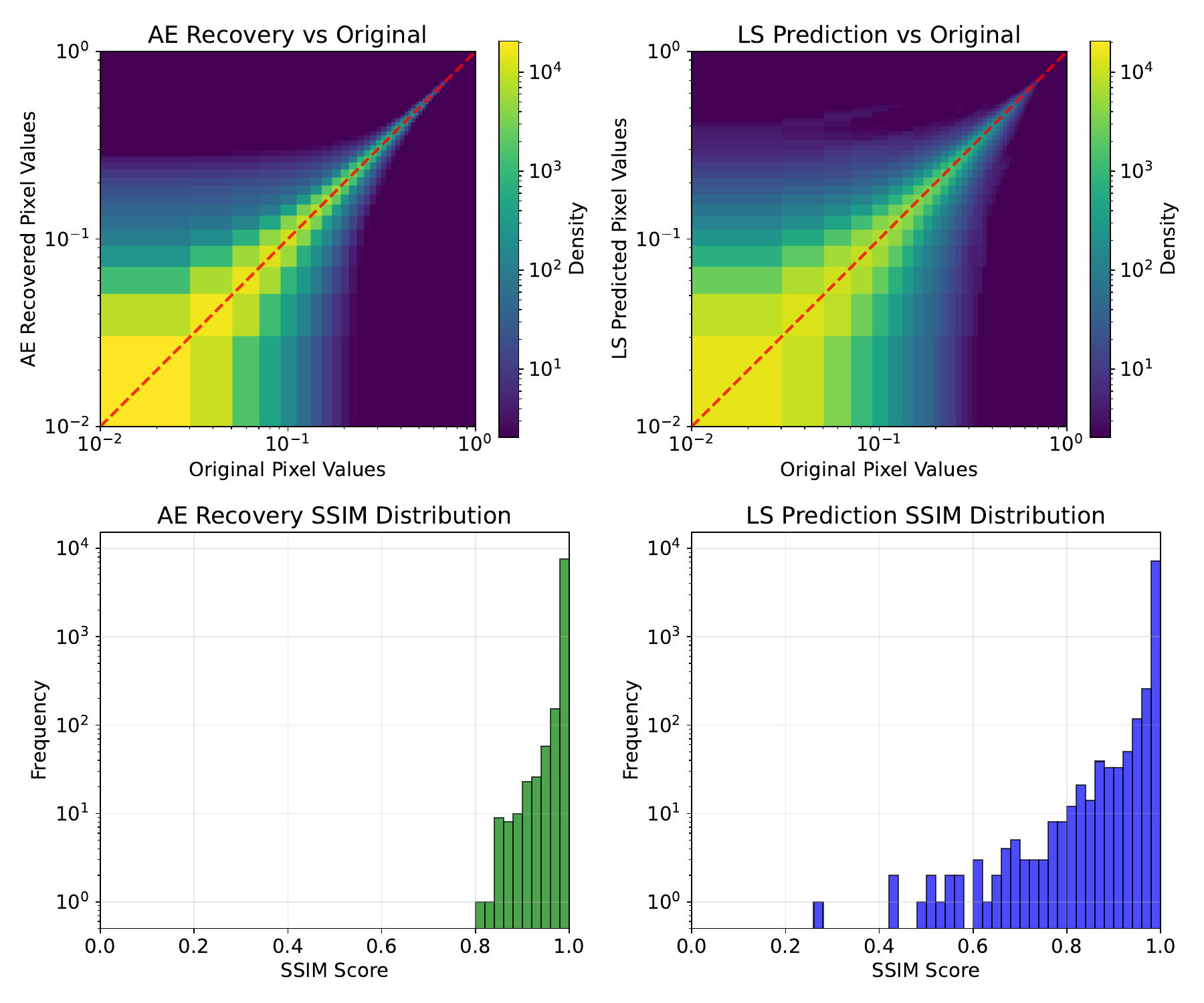}
    \caption{The top panels are two-dimensional histograms comparing the normalised, scaled pixel values between the AE and the validation set (top left) and the latent space decoder model vs the validation set (top right). The bottom two panels show the distribution of the SSIM score of the two neural networks.}
    \label{fig:pixel_comparison_fig}
\end{figure}

In Figure~\ref{fig:pixel_comparison_fig} we show the pixel-by-pixel comparison between the original pixel values of the validation set against pixel values produced by the neural networks. There is excellent agreement, both for the autoencoder images and the images (top left panel) produced from the latent space decoder (top right panel). There is clearly some broadening in the distribution, particularly at the lower brighnesses, but the narrowness of the distribution above a scaled flux of 0.1 suggests that the model images are being very well recovered by the autoencoder. The distribution for the latent space decoder model is broader, as expected, but is still very good.

It is well know that pixel-by-pixel statistical comparisons between images are subject to large variation even when two images are similar. (One could imagine a simple brightness distribution that is offset by a single pixel will have a large mean squared error but would actually look virtually identical to the original). We therefore compared the images using the structural similarity index measure (SSIM, \citealt{Wang2004}), which is a perception-based model that is widely used in signal processing to compare a degraded image $x$ with a distortion-free reference image $y$. 
\begin{equation}
{\rm SSIM}(x,y) = \frac{(2\mu_x \mu_y + c_1)(2\sigma_{xy} + c_2)}{(\mu_x^2 + \mu_y^2 + c_1)(\sigma_x^2 + \sigma_y^2 + c_2)}
\end{equation}
where $\mu_x$ and $\sigma^2_x$ are the pixel sample mean and variance of $x$ (similarly for $y$) and $\sigma_xy$ is the covariance of $x$ and $y$, $c_1 = (0.01 L)^2$, $c_2 = (0.3L)^2$ where $L=2^{32}-1$. The SSIM is calculated over a block window size of $8\times8$ and averaged over the whole image. An SSIM of $+1$ indicates perfect similarity, zero indicates no similarity, and $-1$ corresponds to perfect anticorrelation.

The SSIM distributions for the two networks are plotted as histograms in Figure~\ref{fig:pixel_comparison_fig}. There is a sharp peak maximum in both distributions at an SSIM of $\sim 1$, indicating that the networks are predicting images with close to perfect similarity to the original. The distribution for the latent space decoder images is broader, with a handful of models showing SSIM scores of $<0.6$. 

\begin{figure}
	\includegraphics[width=\columnwidth]{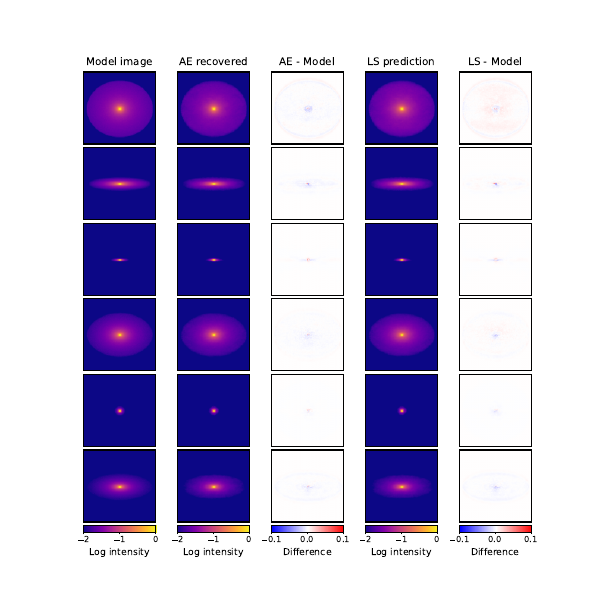}
    \caption{A demonstration of the fidelity of the reproduction of the RT model images using the machine learning methods, using six randomly selected images from the validation dataset. The first column shows the RT model image, the second shows the autoencoder recovered image, and the third shows the linear residuals. The fourth column shows the same model computer from the decoder using the latent space found by applying the NN to the appropriate disc parameters. The final column shows the difference between this and the original model image.}
    \label{fig:validation_images_fig}
\end{figure}

Statistically at least  the networks appear to be capable of reproducing the images. As a final demonstration we plot a random sample of 6 images from the validation set, along with their network reproductions from the AE and from the LS decoder network and their respective residuals (see Figure~\ref{fig:validation_images_fig}). The by-eye comparison is very favourable -- it is hard to distinguish the original images from the network recovered versions. The residuals show that at least some of the pixel-to-pixel deviation is due to the pixel noise in the original images themselves, since these variations are not learned by the autoencoder but are in fact smoothed out by the autoencoder's convolutional layers. There is some evidence that the residuals are larger at he outer radii of the discs, and as expected the residuals are larger for the latent space decoder images. Given the inherent noise in the ALMA data, and the smoothing of the images due to the ALMA beam, the recovered images are perfectly good for fitting.

\section{Determining disc parameters from simultaneous fits to images and SEDs}

In the following section we report a first application of our new method to a homogeneous set of 1.3\,mm images and SEDs in the $\rho$~Oph star-forming region. We first introduce the dataset we are fitting, and then describe how we perform the simultaneous fits. Subsequently we present and comment upon our results.

\subsection{Imaging data}

We used the 1.3\,mm ALMA imaging survey of discs in the $\rho$~Oph cluster obtained as part of the ODISEA campaign \cite{Cieza2019} and kindly provided to the author by L. Cieza and C. Gonzalez Ruilova. These Cycle-4 data were obtained with the C40--5 and 42--45 array configurations of the 12\,m antennae with baselines from 17 to 2647\,m, providing 28-au resolution at the 140\,pc distance to $rho$~Oph. We refer readers to \cite{Cieza2019} for details on the observations and data reduction.

For this study we included all objects with an integrated 1.3\,mm flux of 4\,mJy or greater. (We found that the lowest S/N images didn't provide strong constraints on the disc properties). At this resolution many of the discs are resolved, and we excluded those discs identified by \cite{Cieza2019} as having obvious structure (such as rings) and those discs that are in binary systems. We did however include those systems identified as having hints of structure from analysis of their radial profiles, and we comment on these below. Note that we excluded object 45, a flat spectrum source that is resolved but is structured in a manner that is not consistent with a simple disc model. The final sample comprises 17 Class I objects, 21 flat spectrum sources, and 27 Class II sources.

Each image was first fitted with a 2D gaussian in order to determine the position of the object. The image was then cropped about this central position to $128 \times 128$ pixels (which, given the pixel scale, corresponds to 2.6 arcseconds on a side, or 360\,au at the mean distance of the $\rho$ Oph cluster). The image was then scaled by its maximum value. The pixel noise of the observation was calculated from the standard deviation of the image, excluding the central $64\times 64$ pixels and additionally (for the most extended objects) pixels that likely contained extended emission (determined iteratively by sigma-clipping). 

\subsection{Spectral data}

We constructed an SED for each object primarily using the {\em Spitzer} Cores to Discs catalogue data as provided in Table 2 of \cite{Cieza2019} as well as the 1.3\,mm fluxes. We supplemented these data with $VRI$ band photometry when available in the APASS \citep{Henden2016} or USNO \citep{Monet2003} catalogues, as well as {\em Herschel} photometry from the catalogue of \cite{Rebollido2015}. Since the uncertainties on SED data are often underestimated, we used the maximum of the quoted uncertainty and 10\% of the flux as the uncertainty on each SED point.

\subsection{Constructing the goodness-of-fit quantity}
\label{gof_section}

We use a Bayesian approach to determine the distribution of our disc parameters by a simultaneous fit to the observed SED and the 1.3\,mm images. (We note that in this paper we fitted the data in the reconstructed image plane, but fitting the visibilities should be possible in future).

We adopt a uniform prior ($l(\theta)$) for our vector of model parameters $\theta_i$, bounded by the lower ($\theta_{l,i}$) and upper ($\theta_{u,i}$) bounds of our training data:
\begin{equation}
\log l(\theta_i) = \begin{cases} 0, & \text{if } \theta_{l,i} \leq \theta \leq \theta_{u,i} \\
-\infty, & \text{otherwise.}\end{cases}
\end{equation}

Our goal is to determine a quantity that reflects the likelihood of the fit. Assuming that the SED has $N_{\rm SED}$ flux points ($F_{{\rm obs},i}$) with uncertainties ($\sigma_i$) we define the likelihood of the observations given the model fluxes ($F_{{\rm m},i}(\theta))$ as 
\begin{equation}
L_{\rm SED}(\theta) = \prod_{i=1}^{N_{\rm SED}} \frac{1}{\left( 2\pi \sigma_i^2 \right)^{1/2}} \exp \left( -\frac{1}{2}\frac {(F_{{\rm obs},i} - F_{{\rm m},i}(\theta))^2}{\sigma_i^2} \right),
\end{equation}
or as a log likelihood
\begin{equation}
\log L_{\rm SED}(\theta) = C_1 - \frac{1}{2}\sum_{i=1}^{N_{\rm SED}} \left( \frac {(F_{{\rm obs},i} - F_{{\rm m},i}(\theta))^2}{\sigma_i^2} \right)
\end{equation}
where $C_1$ is a constant. Similarly if the image consists of $N_{\rm pix}$ pixels of intensity  $I_{{\rm obs},i}$ with uncertainties ($\sigma_i$) we can compare the model pixel intensities $I_{{\rm m},i}(\theta)$ to get
\begin{equation}
\log L_{\rm image}(\theta) = C_2 - \frac{1}{2}\sum_{i=1}^{N_{\rm pix}} \left( \frac {(I_{{\rm obs},i} - I_{{\rm m},i}(\theta))^2)}{\sigma_i^2} \right)
\end{equation}
where $C_2$ is another constant. Since we are interested in maximising the likelihood we can neglect $C_1$ and $C_2$ which are constant over parameter space $\theta$.

The separate likelihoods must be combined in order to perform a simultaneous fit. A naive approach would be to declare that the total likelihood $L_{\rm tot}$ is simply the product of the likelihoods of SED and the image, i.e.
\begin{equation}
\log L_{\rm tot}(\theta) = \log L_{\rm SED}(\theta) + \log L_{\rm image}(\theta).
\end{equation}
However for the data considered here $N_{\rm pix} \gg N_{\rm SED}$ which in general means that $L_{\rm image}(\theta) \ll L_{\rm SED}(\theta)$, and the maximum total likelihood value would essentially be completely determined by the imaging data alone.

This so-called data fusion problem is a long-standing one, which is often solved by weighting the contributions of the image and spectral data points in the fitness function (in this case the total likelihood).  Here we chose to weight the imaging data likelihood by a factor $w$, giving
\begin{equation}
\log L_{\rm tot}(\theta) = \log L_{\rm SED}(\theta) + w \log L_{\rm image}(\theta).
\label{eq:ltot_w}
\end{equation}
What value should $w$ take? We will attempt to find a scaling that means that for a reasonable fit $w L_{\rm image}(\theta)$ and $L_{\rm SED}(\theta)$ have a similar magnitude, in order to provide an optimal solution where the imaging and spectral data have a balanced influence on $L_{\rm tot}$ and therefore on the disc parameters. We introduce the z-score
\begin{equation}
z_i  = \frac{I_{{\rm obs},i} - I_{{\rm m},i}(\theta)}{\sigma_i}.
\end{equation}
If we assume that the model is a perfect fit $(\theta=\theta_{\rm p})$ to the data then $z_i$ is  the gaussian random deviate and
\begin{equation}
\log L_{\rm image}(\theta_{\rm p}) \approx -\frac{1}{2}\sum_{i=1}^{N_{\rm pix}}  {z_i^2}.
\end{equation}
The mean variance is 
\begin{equation}
\langle z_i^2 \rangle = 1
\end{equation}
and so
\begin{equation}
\log L_{\rm image}(\theta_{\rm p}) \approx -\frac{N_{\rm pix} }{2}
\end{equation}
with a similar expression for $\log L_{\rm SED}(\theta_{\rm p})$. So we have
\begin{equation} 
w \approx \frac{N_{\rm SED}}{N_{\rm pix}}.
\end{equation}
For the observations here $N_{\rm pix} = 128 \times 128$ and $N_{\rm SED} \sim 10$, and so we take $w = 0.0006$. 

\section{Performing the fit}

For a particular set of disc parameters we ran the two networks to recover the appropriate model SED and image. We first rotated the model image by the position angle (PA, a free parameter of the model). We then resampled the image to match the pixel scale of the observation. The model image size $128\times 128$ pixels and is always 10\% larger than twice the outer disc radius, and the resampling needs to be done carefully, since any resampling that creates pixel-scale artifacts or discontinuities adversely affects the fitting process. We first determined the pixel scale of the model in au/pixel, and converted this to arcseconds/pixel using the appropriate distance. We then resampled the model image onto the the observational scale using a bicubic spline fit to the model pixel intensities. If the resampling made the image significantly smaller (by a factor of greater than 1.5) then we sub-sampled each pixel to ensure flux conservation in the resampled image and to reduce aliasing. The ALMA beam, taken from the FITS headers of the image, was then convolved with the model image using a fast fourier transform method. 

The model SED (which has fluxes that assume a distance of 100\,pc) was first scaled to the appropriate object distance using the GAIA DR3 \citep{Vallenari2023} distance to the object (if available) or the mean distance to the cluster (139.4\,pc) otherwise. We used the {\tt dust\_extinction} python package to redden the spectrum \citep{Gordon2024}. The $A_V$ is a free parameter of the fit, and the extinction law applied was that of \cite{Gordon2023}, which is valid out to 30\,$\mu$m (beyond this wavelength the extinction was extrapolated to zero using log-log extrapolation from the gradient of the extinction law at 30\,$\mu$m). The fluxes at the observed SED wavelengths were determined using log-log interpolation on the dense model wavelength grid.

The log likelihood values, and associated reduced chi-squared values, then calculated using the process described in Section~\ref{gof_section}. (The reduced chi-squared values are used as illustrative of the goodness of fit, and are not employed in the optimisation process. Note that for a small number of objects the number of observations in the SED was fewer than the number of fitted parameters, and the number of degrees of freedom for the SED reduced chi-squared was set to unity in these cases).

There is a considerable parameter hyperspace to search, and it is dotted with local maxima in likelihood. We adopted a hybrid optimisation scheme to identify the region of global maximum likelihood. First we used a genetic algorithm (as implemented by \citealt{Gad2023}), which searched the entirety of valid parameter space, using 2000 models per generation and 200 parents breeding, and applying a steady state selection algorithm with a 10\% mutation rate. After 20 generations this algorithm has typically converged to a satisfactory fit, and the most likely model in that final generation used as a starting point for a downhill simplex algorithm using the Nealder-Mead method \citep{Nelder1965} as implemented in {\tt scipy}, which typically converges after 2000 steps. 

Finally we used the Markov Chain Monte Carlo (MCMC) method (via the python module {\tt emcee} by \citealt{emcee2013}) in order to calculate the posterior distribution of the parameters, using 50 walkers of 10\,000 steps each. The starting point of the walkers was the downhill simplex solution, perturbed by small gaussian random deviates, and we conservatively removed the first 10\% of the MCMC chains as burn in. We constructed corner plots in order to visualise the solution, check for anomalies in the fits, and determine the confidence intervals on the results.

It is worth noting that identifying a solution for a single object, from the genetic algorithm, through the downhill simplex, and using the MCMC algorithm, typically requires the generation of over half a million model SEDs and images. This would be wholly impractical without the machine learning method.

\subsection{Fitting the data}

Although the disc model is best suited to the Class II sources in the sample, we apply it to both the Class I and flat spectrum sources as well. It is certainly possible that at least some of the flat spectrum sources are Class II objects but viewed in an edge-on configuration, obscuring direct stellar radiation, whereas others may represent a transition stage between Class I and Class II. The model's applicability to the more embedded Class~I objects is undoubtedly more questionable, but there are reasons for optimism: The high-resolution ODISEA observations spatially filter-out structures on the order of 1000\,au or more, meaning that the 1.3\,mm flux is from the disc itself and the image comparison is therefore valid. Spectrally the model presented here neglects the obscuration of the source and disc emission by the extended envelope of the Class I source, and more importantly the star/disc emission that is re-processed by the  envelope. The former can be simulated by allowing a very high reddening (although the extinction may not be spectrally the same as an envelope), the latter is a flaw that cannot be easily mitigated without running a bespoke training set and accounting for the ALMA spatial filtering by fitting to the visibilities as opposed to the reconstructed image. In summary the results should be treated with increasing caution as we progress from Class II objects, through the flat spectrum sources, to Class I objects.

The modelling process allows us to fit for 11 or 12 free parameters: the dust mass $M_{\rm dust}$, the inner and outer disc radius ($r_{\rm in}$ and $r_{\rm out}$, the flaring index $\beta$, the surface density power law index $\gamma$, the $h/r$, the dust settling $\xi$, the inclination $i$, the PA of the disc, the stellar radius $R_*$, and the reddening $A_V$.  For objects without spectral types we additionally fitted for $T_{\rm eff}$. When spectral types are available we used the spectral type--$T_{\rm eff}$ calibration of \cite{Pecaut2013} to fix the effective temperatures.

\section{Results}

\begin{figure*}
	\includegraphics[width=180mm]{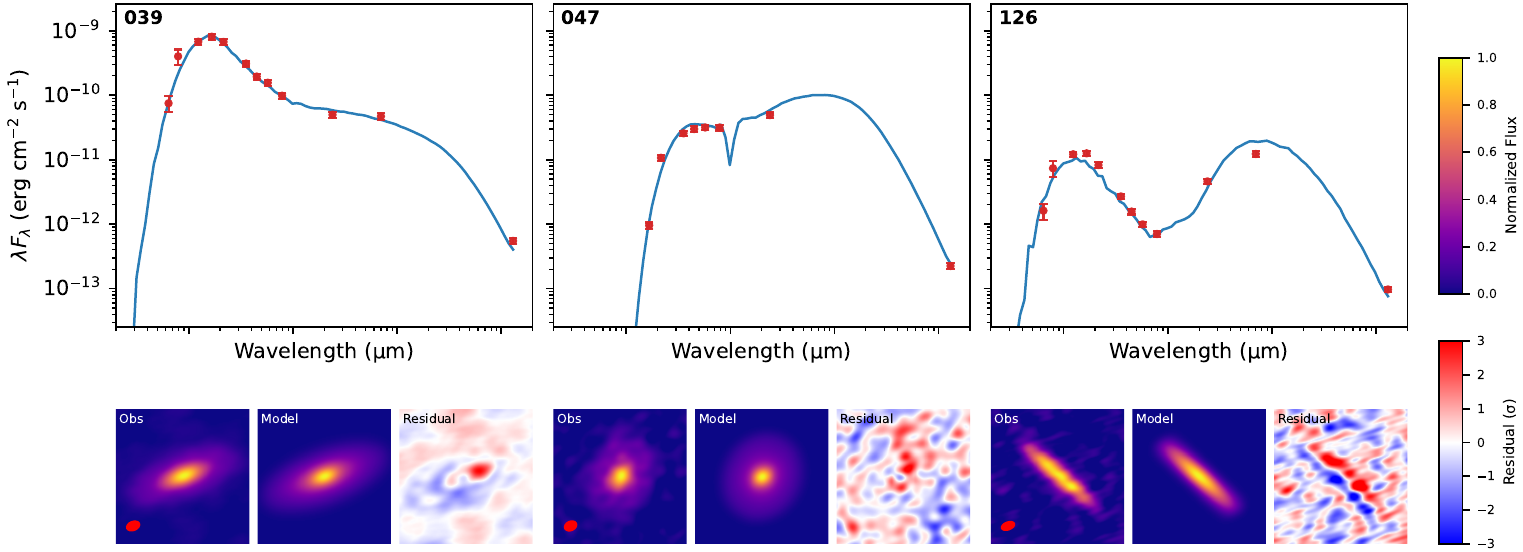}
    \caption{Example fits for a Class II source (ID 039), a Class I source (047), and a flat spectrum source (126). The observed SED (red symbols) are plotted with the best fit model SED (blue line). The lower panels show the normalised observed 1.3 mm image, the best fit model, and the residuals (observed minus model). The images are 2.6 arcsecond on a side.}
    \label{fig:fit_examples}
\end{figure*}

We plot examples of the best fits for resolved Class II, Class I, and flat spectrum sources in Figure~\ref{fig:fit_examples}. One can see that the SED fit is generally satisfactory, although the lack of IR datapoints means that the long-wavelength part of the SED is not strongly constrained. The imaging fits are generally quite good, although the example objects still have significant residuals.

\begin{figure*}
	\includegraphics[width=180mm]{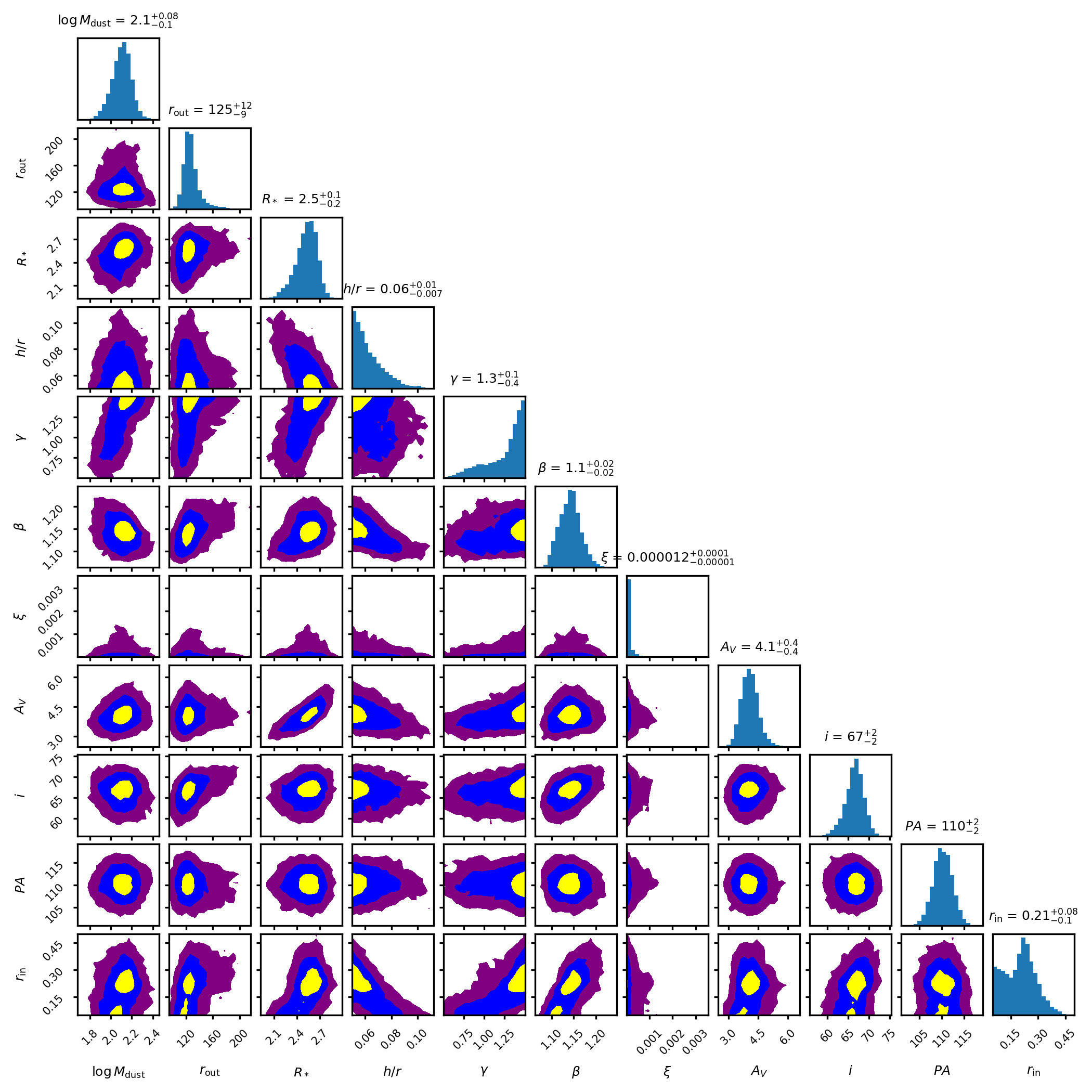}
    \caption{The corner plot for Class II object 039. The histograms on the diagonal represent the marginalized posterior densities for each free parameter. The off-diagonal scatter plots show two-dimensional projections of the samples to reveal covariances. The yellow, blue, and purple colours represent the $1\sigma$, $2\sigma$, and $3\sigma$ confidence intervals respectively. }
    \label{fig:corner_example}
\end{figure*}

We give an example corner plot (for the Class II object 039) in Figure~\ref{fig:corner_example}. The success of the fit leads to quite well defined peaks in posterior distribution, in particular parameters such as the outer radius, the inclination and position angle are all strong constrained by the well-resolved imaging data. However one can also identify some of the limitations of the approach. For example the radial surface density power law index ($\gamma$) and the $h/r$ are being constrained by the upper and lower limits of the prior distribution. The inner radius of the disc $r_{\rm in}$ also shows that solutions close to the sublimation radius are good, as are solutions with a reasonably substantial (0.2\,au) inner hole to the disc.

The SED and imaging fits for the entire dataset are presented in Figure~\ref{fig:full_plot} with the best-fit parameters and their uncertainties listed in Table~\ref{tab:full_results}. The imaging fits shown in Figure \ref{fig:full_plot} are typically good, and this is confirmed quantitatively, with the fits yielding a reduced $\chi^2$ of close to unity (Table~\ref{tab:chisq}) across all spectral classes. The SED fits are rather less successful, with the Class II objects having a median reduced $\chi^2$ of $\sim 7$, followed by a decline in the quality of the SED fits for flat spectrum sources ($\sim11$) and finally Class I objects having reduced $\chi^2 \sim 19$. Some of this is due to underestimates of the true random and systematic uncertainties on the photometric observations, combined with inherent variability that is present particularly in the optical and IR. However inadequacies of the underlying model certainly contribute, in particular the absence of the envelope for the Class I and flat spectrum sources.

\begin{table}
\caption{Median reduced $\chi^2$ values for the SED and 1.3\,mm image fits, broken down by class.}
\label{tab:chisq}
\begin{tabular}{llll}
\hline
Class & $N$ & $\overline{\chi^2}_{SED}$ & $\overline{\chi^2}_{\rm image}$ \\
\hline
I & 17 & 18.9 & 1.7 \\
F & 21 & 10.8 & 1.2 \\
II & 27 & 6.9 & 1.5 \\
\hline
\end{tabular}
\end{table}

\begin{figure*}
	\includegraphics[width=180mm]{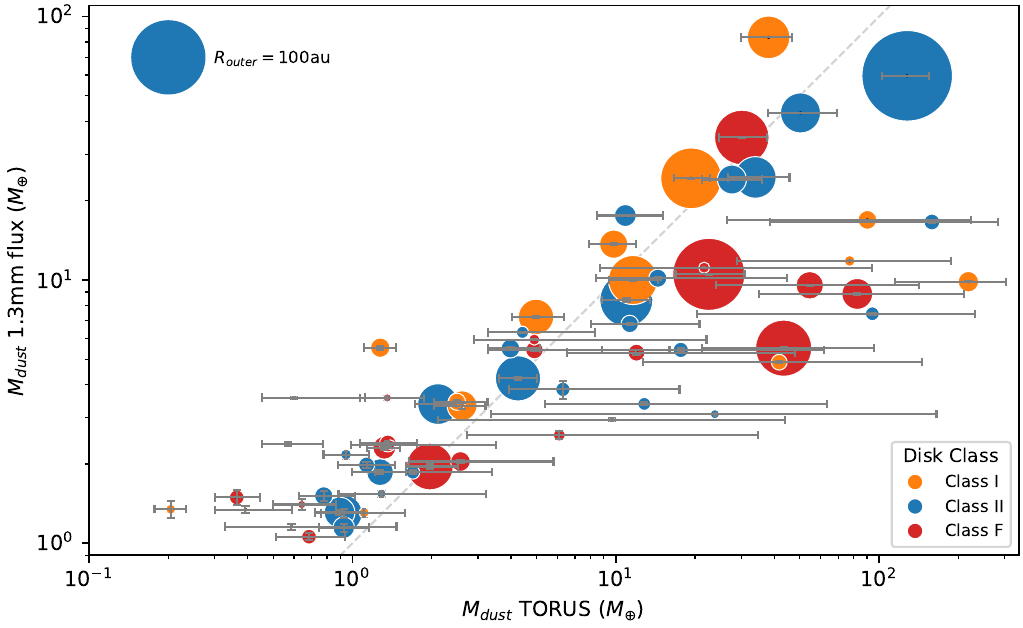}
    \caption{The 1.33\,mm sub-mm flux dust mass estimates plotted against those found using objective fitting of the SEDs and images. The uncertainties are derived from the uncertainty on the 1.33\,mm fluxes (vertical) or from MCMC estimates (horizontal). The radius of the disc is illustrated by the size of the symbols, and an example for a 100\,au-radius disc is given at the top left. The dashed grey lines show the expected 1:1 relationship.}
    \label{fig:mass_plot}
\end{figure*}

\section{Discussion}

A comparison of the $M_\mathrm{dust, 1.3\,mm}$ estimates and those derived from the objective fits to the SED and images ($M_{{\rm dust}, {\sc torus}}$) reveals some interesting effects that are worthy of more detailed discussion. We plot the two dust mass estimates against each other in Figure~\ref{fig:mass_plot}, distinguishing between the SED classifications by symbol color and the disc radius by symbol size.

One can immediately see that the objects, to first order, follow the 1:1 line, an endorsement of the use of the 1.3\,mm flux as a dust mass proxy by \cite{Williams2019}. It is also apparent, also to first order, that Class I and II sources (the orange and blue symbols in Figure~\ref{fig:mass_plot}) follow the 1:1 with similar scatter. The flat spectrum sources (red symbols) however show significant deviations away from the 1:1 line, with a cluster of larger discs with dust masses of 10 to 100\,M$_{\earth}$ that fall well to the right of the 1:1 line, indicating that they have higher masses than their 1.3\,mm fluxes would suggest. Indeed these are objects (for example IDs 26, 44, 72, 64 126) that are likely Class II sources viewed at high inclination where optical depth effects are maximised (this is discussed further below).

Another obvious feature of Figure~\ref{fig:mass_plot} is how poorly the dust mass is constrained for many of the objects (compare the size of the horizontal error bars to the vertical ones). The uncertainties are largest for objects that lie to the right of the 1:1 line, meaning that these are objects for which the 1.3\,mm dust masses are under-estimates, and we discuss this further below. One can also see strong evidence of a disc radius/mass correlation, with the high mass objects dominated by systems with radii of $>50$\,au.

Focussing on the low-mass regime a broadly linear relationship between the two mass estimates does exist, but with significant scatter. We note that for the majority of low-mass objects fall  above the 1:1 line (i.e. $M_\mathrm{dust, 1.3\,mm}$ is over estimated by a factor of a few at low masses). Given the disc-mass--radius relationship revealed by the sub-mm imaging, the principle reason for this discrepancy is the dust temperature, which for smaller discs is significantly greater than 20\,K assumed in deriving $M_\mathrm{dust, 1.3\,mm}$. 

We ran {\sc torus} models using the fiducial parameters list in Table~\ref{tab:fiducial}, calculating the mass-weighted average mm-dust temperatures for a 10\,M$_{\earth}$ dust mass disc with radii of 20\,au, 50\,au, 100\,au and 200\,au, which encompasses the range of disc sizes that we see in the ODISEA sample. Unsurprisingly the smaller, more strongly irradiated discs have higher dust temperatures. The 200\,au disc has mean mm-dust temperature of 27\,K, increasing to 34\,K ($r_{\rm out}=100$\,au), 43\,K (50\,au) and finally 61\,K (20\,au). Thus we expect the smaller discs, that occur typically (but not exclusively) at low masses to have dust masses that our overesimated by the 1.3\,mm flux, as observed in Figure~\ref{fig:mass_plot}. Conversely we note that the agreement between $M_\mathrm{dust, 1.3\,mm}$ and $M_{{\rm dust}, {\sc torus}}$ improves for the resolved Class~II discs, since the dust-mass-weighted temperatures of grains in these larger discs are  typically 20--30\,K, and thus in better agreement with the assumption underpinning the sub-mm dust mass estimate.

\begin{table}
\caption{Parameters of the fiducial disc model.}
\label{tab:fiducial}
    \begin{tabular}{lll}
    \hline
    Parameter & Value  & Unit\\
    \hline
    $T_{\rm eff}$ & 4000 &K \\
    $R_*$ & 2 &$R_\odot$ \\
    $\beta$ & 1.125 &\\
    $h/r$   & 0.1 & \\
    $\gamma$ & 1 & \\
    $r_{\rm in}$ & 0.06 &au \\
    $\xi$ & 0.1  & \\
    \hline
\end{tabular}
\end{table}

We can examine the optical depths and dust temperature on the 1.3\,mm flux in more detail by using the SED neural network. We employ the same fiducial disc, whose parameters are given in Table~\ref{tab:fiducial}. We used the SED network to predict the 1.3\,mm flux, and hence the corresponding dust mass estimate using Equation~\ref{eq:mdust_flux}, for a range of dust masses, disc outer radii, and inclinations.

\begin{figure}
	\includegraphics[width=88mm]{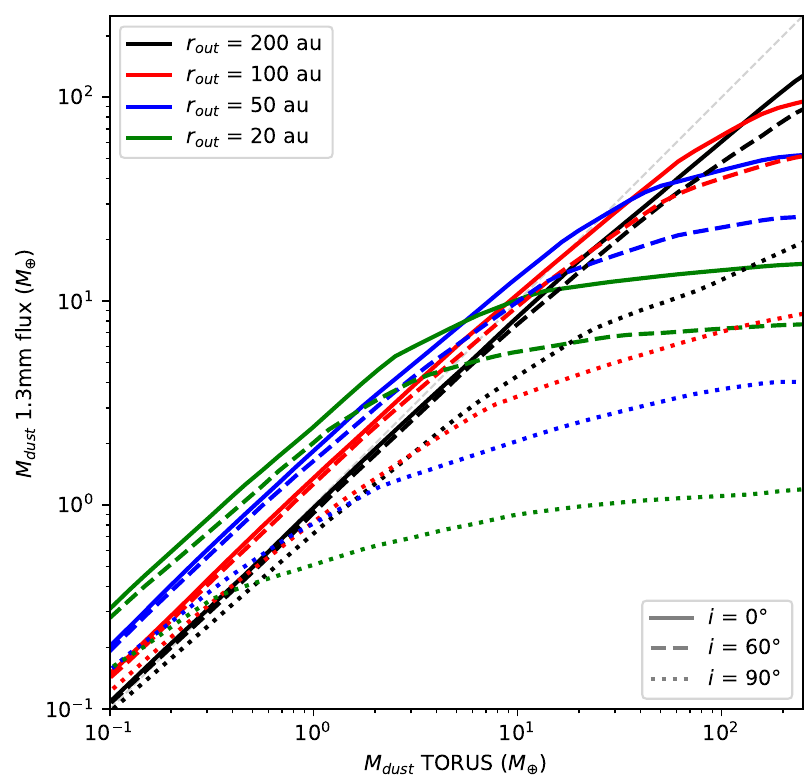}
    \caption{The 1.33\,mm sub-mm flux dust mass estimates computed from the fluxes predicted by the detailed RT modelling using the fiducial disc model described in the text. The different coloured lines refer to different outer disc radii, with the different line styles corresponding to different system inclinations.}
    \label{fig:mass_plot_2}
\end{figure}

The results of these calculations are plotted in Figure~\ref{fig:mass_plot_2}. For the largest discs  ($r_{\rm out}=200$\,au, black lines) viewed face on ($i=0^\circ$, solid linestyle) the 1.3\,mm dust mass estimate is good, tracking the 1:1 relationship at low masses and only deviating for dust masses of greater than approximately 10\,M$_{\earth}$. For these discs the bulk of the midplane dust is close to 20\,K and the face-on orientation, combined with the larger radius, means that optical depth effects are minimised. Optical depth effects do come into play as the $r_{\rm out}=200$\,au discs' inclination increases to $60^\circ$ (black dashed line) and then $90^\circ$ (dotted black line), with the highest mass discs having significantly higher masses than predicted by the 1.3\,mm flux. Reducing $r_{\rm out}$ to 100\,au follows a similar pattern, but with the low mass discs having their mass slightly overestimated by the 1.3\,mm flux, and the higher mass discs being more sensitive to optical depth effects as the disc material is hotter and confined to a smaller volume. This pattern continues as we move to smaller and smaller discs. For the smallest discs ($r_{\rm out}=20$\,au, green lines) the dust masses are systematically overestimated at low masses by a factor of a few, and underestimated at high masses by an order of magnitude or more depending on inclination. 

Figure~\ref{fig:mass_plot_2} also demonstrates how, particularly for the smaller discs, the 1.3\,mm flux loses its diagnostic power at higher masses, leading the large uncertainties on the dust masses found during the fitting process.

\begin{figure}
    \includegraphics[width=88mm]{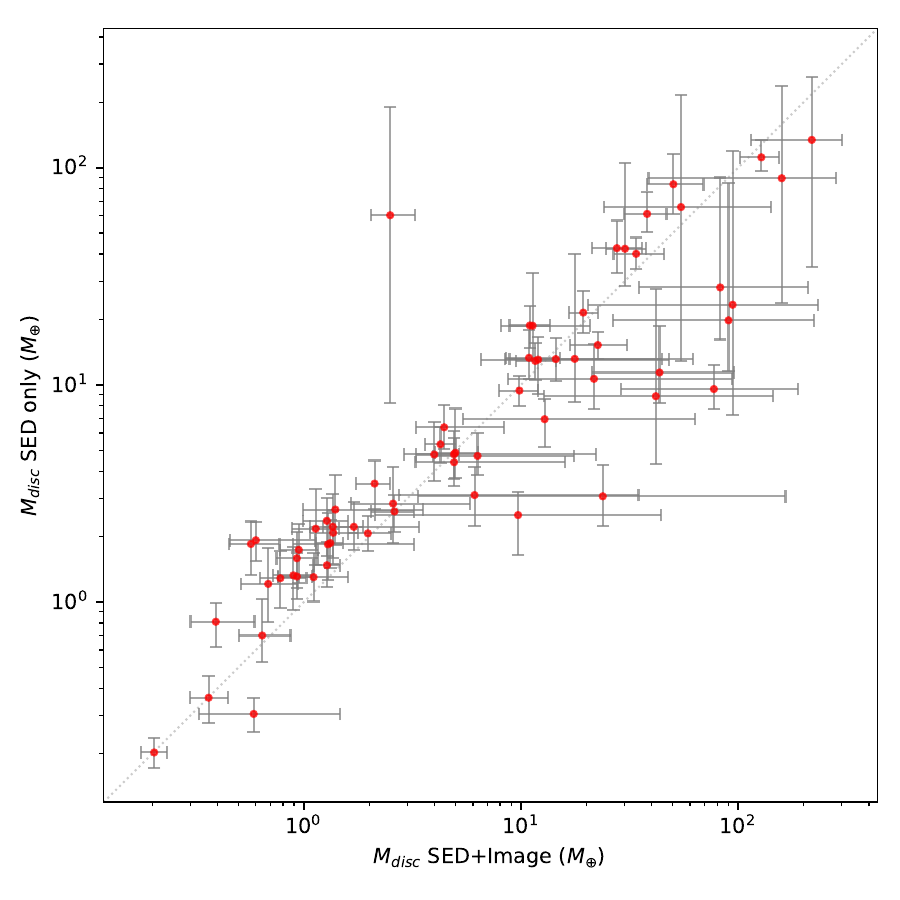}
    \caption{A comparison between dust mass derived using the SED along versus the dust mass derived from simultaneous fits to the image and the SED.}
    \label{fig:mass_comp}
\end{figure}

Often the imaging data is not available, and studies fit the SEDs alone without any size constraint (e.g. \citealt{Ribas2020}). Given that it appears that disc sizes seem critical, one might question how well these fits recover the dust mass? In order to investigate this we ran the full multi-dimensional optimisation for the sample but on the SED alone (essentially setting the imaging weight factor $w=0$ in Equation~\ref{eq:ltot_w}) and plotted the dust masses derived against the results from the simultaneous fits (Figure~\ref{fig:mass_comp}).

The broad tendency for the SED-only fit is to converge to solutions with a large disc radius that would be inconsistent with the imaging data. These solutions have a slightly higher disc mass (the mean grain temperatures will be lower so more dust is needed in the model to fit the sub-mm point). The results are encouraging for SED-only studies, since  the masses are scattered around the 1:1 line, albeit with a tendency for the SED-only  models to slightly over-estimate the mass. There is a however a large subset of objects (primarily those that are strongly constrained to small radii by the imaging data for the simultaneous fits) for which the dust masses are severely underestimated by the SED-only fit. 

\begin{figure}
	\includegraphics[width=88mm]{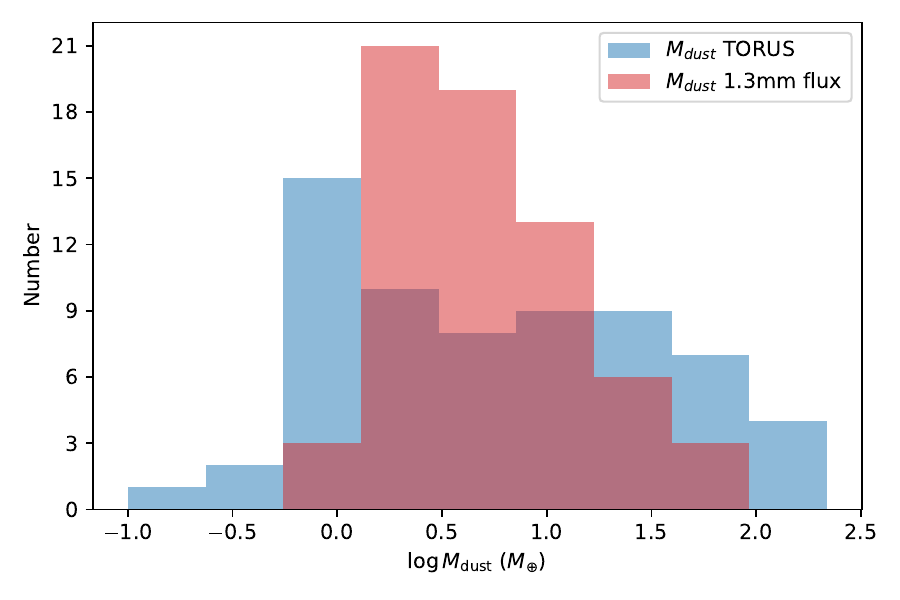}
    \caption{Histograms of dust mass derived from the 1.3\,mm estimate (pink) and the {\sc torus} models (blue). }
    \label{fig:mass_hist}
\end{figure}

In Figure~\ref{fig:mass_hist} we compare the dust masses derived from the 1.3\,mm fluxes with those from the SED/image fitting. It can be seen that the {\sc torus} dust mass distribution is broader, with an excess of objects at the higher-mass end of the distribution and at the lower-mass end. The distribution is therefore less peaked, with the peak distribution occuring at lower mass. Note that the median of log dust mass marginally {\em increases} for the {\sc torus} distribution, but the effect of the broadening is quite substantial: 13/65 (20\%) of objects in the {\sc torus} sample have dust masses less than 1\,M$_{\earth}$, whereas there are none in the 1.3\,mm estimates. Similarly  27/65 (42\%) of the objects in the {\sc torus} distribution have masses greater than 10\,M$_{\earth}$, which is nearly double that found from the 1.3\,mm flux (15/65, 23\%). For the most massive discs the {\sc torus} sample has 3 discs with a mass greater than 100\,M$_{\earth}$, whereas the 1.3\,mm sample has none.

\begin{figure}
	\includegraphics[width=88mm]{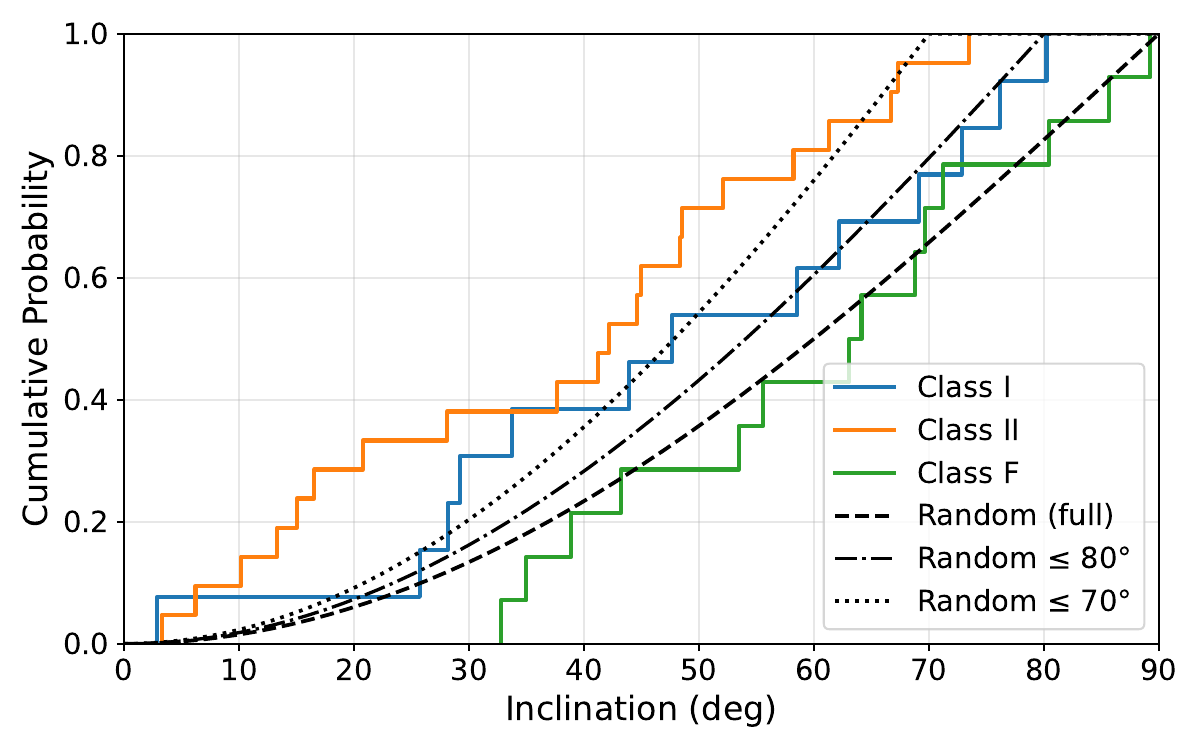}
    \caption{Cumulative probability distributions for the inclinations of resolved discs, broken down by spectral class. The expectation for random inclinations is also shown (dash line) along with random inclinations truncated at $70^\circ$ (dotted line) and $80^\circ$ (dot-dashed line).}
    \label{fig:cdf_inclination}
\end{figure}

We examined the distribution of inclinations for the resolved discs, broken down by spectral class (Figure~\ref{fig:cdf_inclination}). A KS test shows that both the Class I and flat spectrum sources are consistent with random inclinations. In this dataset therefore there is, perhaps surprisingly, not strong evidence that the flat spectrum sources are biased to edge-on systems. However, the CDF for the Class II sources is not consistent with a random distribution ($p=0.035$). We therefore checked the distributions against a random distribution where objects have a maximum inclination of $80^\circ$--this too is inconsistent with the observations ($p=0.031$). The CDF is however consistent with a maximum inclination of $70^\circ$. This is on the face of it an expected result, since below $\sim 70^\circ$ there will be significant effects on the SED as the direct stellar radiation begins to be obscured by the disc.

\begin{figure}
	\includegraphics[width=88mm]{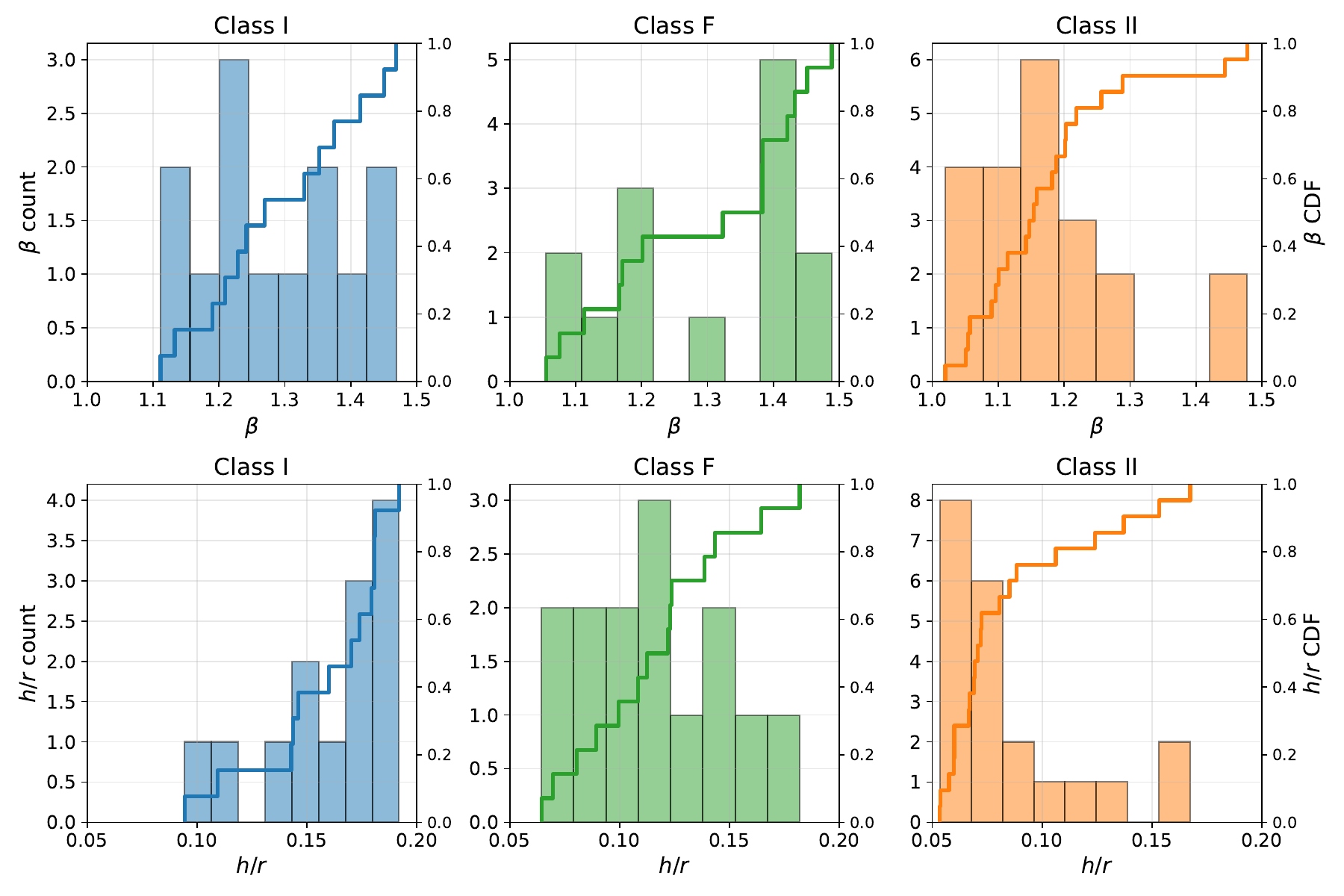}
    \caption{Cumulative probability distributions (lines) and histograms of the distributions of flaring parameter $\beta$ (top row) and $h/r$ (bottom row) broken down by spectral class.}
    \label{fig:beta_hr_distributions}
\end{figure}

We explored the distributions of the disc flaring power-law index $\beta$ and disc scale height $h/r$ broken down by spectral class (see Figure~\ref{fig:beta_hr_distributions}). Focusing first on $\beta$ we find that a KS tests reveals that the distributions for Class I and flat spectrum sources are indistinguishable. However the Class II distribution is statistically different to the Class I and flat spectrum sources, with a greater number of small values (less flared) discs. Looking at the $h/r$ distributions, the KS test shows that the Class I, flat spectrum, and Class II distributions are mutually inconsistent, with $h/r$ values peaking at larger values for the Class I sources, but a much flatter distribution for the flat spectrum sources. The Class II sources show an $h/r$ distribution that is peaked at low values.

\section{Conclusions}

Recent studies have shown how it is possible to use neutral networks to speed up what is effectively detailed radiative-transfer modelling of the discs (e.g. \citealt{Ribas2020}), meaning that studies of large number of objects is possible. However the SED alone does not provide sufficient constraints on the discs, with the situation being ameliorated by adding size constraints \citep{Ballering2019} from thermal dust images. In this study we have taken a further step towards multi-messenger fitting of protoplanetary discs, simultaneously fitting both the SED and the imaging data by using neural networks to compute model images and spectra.

Our results show that this detailed modelling is important for our understanding for dust evolution in protoplanetary discs. \cite{Williams2019} put forward cogent and powerful arguments for the use of the 1.3\,mm dust mass estimate, and indeed our results provide support for their position (see Figure~\ref{fig:mass_plot}). However we need to recognise the effect the outliers have: The mass distribution is significantly revised (see Figure~\ref{fig:mass_comp}), leading to an increase in objects with both massive and low-mass discs. This is not the uniform scaling of the dust mass that occurs when changing the opacity in Equation~\ref{eq:mdust_flux}, nor do the changes arise from a scaling of the dust temperature with stellar effective temperature. Instead it is comes from placing individual discs in their astrophysical context, by combining the SEDs and the imaging. The imaging of course puts tight constraints on the physical size of the disc, but it is also very helpful in constraining the inclination (critical of controlling for optical depth effects), to a lesser extent, the radial power-law of the discs, and their heights and flaring. We have previously pointed to the limitations of the training set here, and particularly in its application to the Class I sources, and to a certain extent the flat spectrum objects. Improving this model should be a priority for the future.

From a technical standpoint the machine learning aspect of this research has fared well. The image reconstruction is generally good, but the obvious degradation of the image reproduction when comparing the autoencoder with the latent space/decoder network, particularly when examining the SSIM histograms, (see for example Figure~\ref{fig:pixel_comparison_fig}, and Figures~\ref{fig:ssim_grid} and \ref{fig:ls_grid_row}) merits further discussion. As mentioned in the introduction, the latent space can often be disordered, and simple interpolation in this space in the hope of reproducing meaningful images can be fraught with difficulty. The fact is though that this method works well for this particular problem (perhaps because the morphology of the thermal dust images is relatively simple and the variation of that morphology with disc parameters is generally slow and predictable). There are methods available to bring more order to the latent space, which will be worth exploring in future. For example one could adopt of some aspects of the variational encoder (VAE)  \citep{Kingma2014}, which uses the Kullback-Leibler divergence \citep{Kullback1951}  as a penalty function to regularise the latent space and ensure that it is ordered, smooth and continuous. VAEs have a latent space that is described by pairs of means and variances, and allows a smooth interpolation in the latent space in order to create novel images that have some qualities that are combinations of features of images in the training data.

The success of the new method opens up several avenues for future exploration. The most obvious stemming from this study is that an extended training set, incorporating a disc plus envelope model, with polar cavities, would allow more reliable fits to the Class I and flat spectrum sources in the ODISEA sample. This would probably necessiate an additional step to account for the spatial filtering of the sub-mm flux during the calculation of the goodness-of-fit.

Secondly, it is possible to calculate images at multiple wavelengths for each disc model in the training set, and therefore perform in-depth studies of objects that have multi-wavelength data. This is  a similar approach to that adopted by \cite{Pinte2008} for IM~Lup, but without the necessity to calculate a bespoke grid for each object. Futhermore by pushing up top of the stellar mass range in the training set one can start to model the extensive datasets for YSOs (for example \citealt{Lumsden2013}),  extending the grid-based approach pioneered by \cite{Robitaille2007}.

Although the current method is focused on imaging data, it will certainly possible to extend the method to line data, either modelling spectral datacubes themselves or the moment maps. This would enable the modelling both the thermal dust and the molecular emission simultanteously, an important step in understanding the formation of planetary systems.

Finally, as higher-resolution ALMA images become available, it is becoming clear that structures within discs are ubiquitous, in particular gaps and rings of dust, which may or may not coincide with gaps and rings of gas. The training model here is based on a smooth, continuous disc structure, and does not account for unresolved structure in the discs. There are two ways forward: The first is to extend the training set to include gaps and rings - this is a formidable task given the wide-range of likely disc configurations. Nonetheless, giving sufficient computing time and a robust statistical model for the likely distribution of discs structures (numbers of rings, size of gaps etc) it would be possible to include such discontinuities in the training set. 

Another approach would be to use the power of the autoencoder to try and identify the presence of disc structures in poorly resolved data. The latent space distribution encodes the expectation for the morphology of different discs in the training set. Simply put, if an image of an unstructured disc is passed into the autoencoder, the encoding  will result in a latent space vector that is consistent with the distribution of latent space vectors used in the training. Alternatively, if the latent space vector of a particular disc lies outside this distribution it may be labelled as anomalous, and one may infer that it will have structures within it (such as gaps or rings) that are not present in the training set.

\section*{Acknowledgements}

I give my warmest thanks to Lucas Cieza and Camilo Gonzalez Ruilova, and the rest of the ODISEA consortium, for providing the 28\,au, 1.3\,mm survey images. I am very grateful to Matthew Bate, Sebaastian Krijt, and Seba Marino for useful discussions. This paper makes use of the following ALMA data: ADS/JAO.ALMA\#2016.1.00545.S. ALMA is a partnership of ESO (representing its member states), NSF (USA) and NINS (Japan), together with NRC (Canada), NSTC and ASIAA (Taiwan), and KASI (Republic of Korea), in cooperation with the Republic of Chile. The Joint ALMA Observatory is operated by ESO, AUI/NRAO and NAOJ. This work made use of Astropy:\footnote{http://www.astropy.org} a community-developed core Python package and an ecosystem of tools and resources for astronomy \citep{astropy:2013, astropy:2018, astropy:2022}. The training set was computed on the University of Exeter's high performance computing system ISCA. 

\section*{Data Availability}

The imaging data presented in this paper are part of the ODISEA survey, and are available in the ALMA archive ({\tt https://almascience.eso.org/aq/}).



\bibliographystyle{mnras}
\bibliography{paper} 




\appendix

This is the appendix.

\section{Optimising the hyperparameters of the neural networks}
\label{appendix_a}

In this section we describe the methods we used to optimise the three neural networks employed, i.e. the SED generator, the autoencoder, and finally the latent space/decoder network.

\subsection{The SED generator}

Following the findings of \cite{Ribas2020} we adopted a network with two hidden layers connecting the input parameters to the 100 SED flux outputs, as illustrated in Figure~\ref{fig:sed_nn}. We used a batch size of 64, with the Adam optimiser and the loss function was selected to be the mean absolute error (MAE). We explored sizes of 10, 20, 50, 100, 200, 500, and 1000 neurons per layer.

\begin{figure}
    \includegraphics[width=88mm]{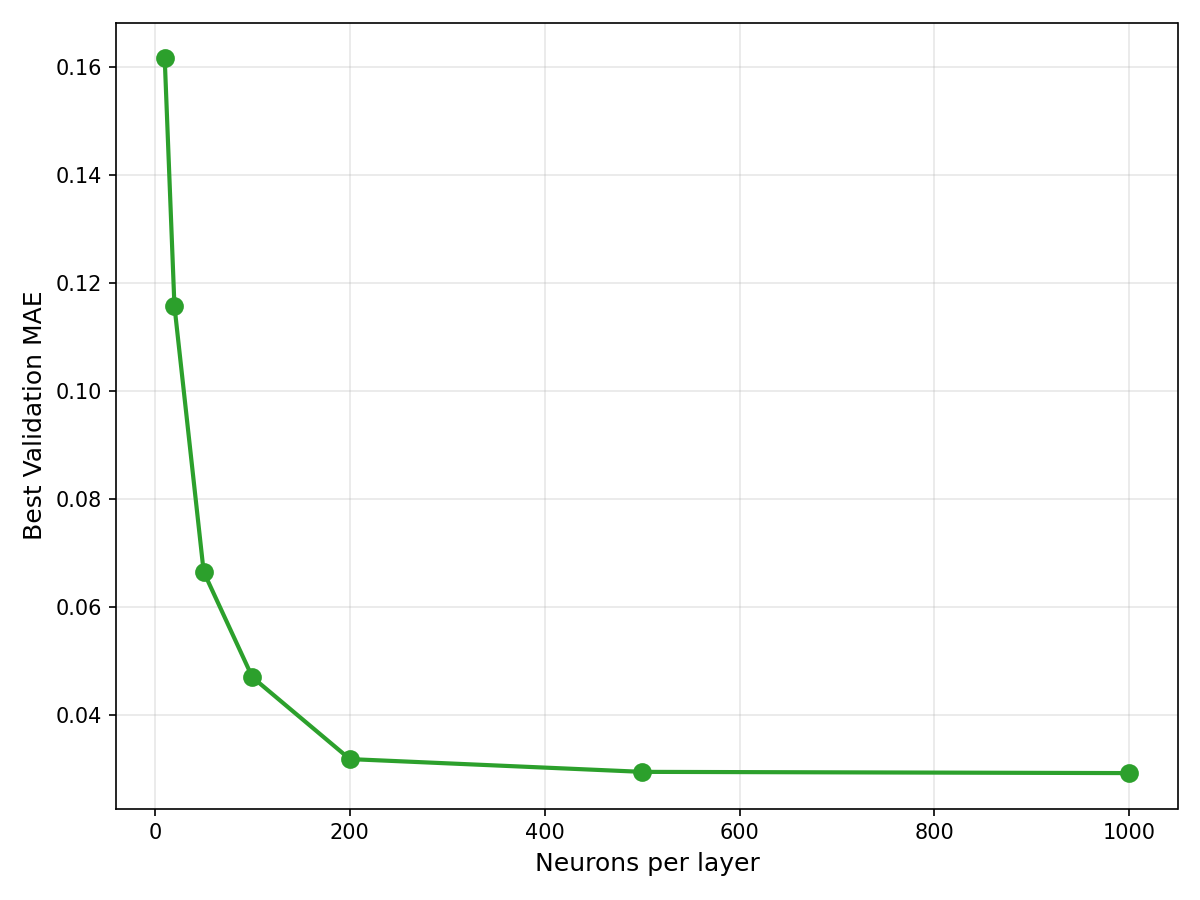}
    \caption{The best validation loss for the SED generation network plotted against the number of neurons per layer.}
    \label{fig:sed_optimisation}
\end{figure}

The validation loss is plotted against neurons per layer in Figure~\ref{fig:sed_optimisation}. It is immediately apparent that the validation loss function decreases rapidly with the number of neurons per layer as expected (the number of free parameters in the network is growing rapidly). The minimum MAE occurs for the 500 neurons per hidden layer network, but the improvement in the loss function is just 8\%, despite increase in the number of free parameters. We note that the validation loss for the 500 neuron network plateaus at early epochs during the training, while the training loss continues to decrease, which hints that we are starting to see an element of over-fitting. By 1000 neurons per layer the validation loss is higher than the training loss, a strong indication of over-fitting. We therefore select the 200 neuron per hidden layer configuration, similar to the \cite{Ribas2020} network (which contained 250 neurons per layer).

\subsection{The autoencoder}

We selected a configuration in which the base units were convolutional layers of with $3 \times 3$ filters and a stride of two, reducing the size of the images by a factor of two each time. Starting with the original $128 \times 128$ pixel image, this reduces the original to a size of $16 \times 16$ after three convolutional layers. The output of each convolutional layer is a number ($N_{\rm filter})$ learned patch filters that result in $N_{\rm filter}$ feature maps. The output of the final layer is therefore a $16 \times 16 \times N_{\rm filter}$ tensor, which is flattened and fully connected to the latent space vector (of size $L_N$). This process is then reversed using convolutional transpose layers to reproduce the input image (see Figure~\ref{fig:autoencoder}). 

\begin{figure*}
    \includegraphics[width=180mm]{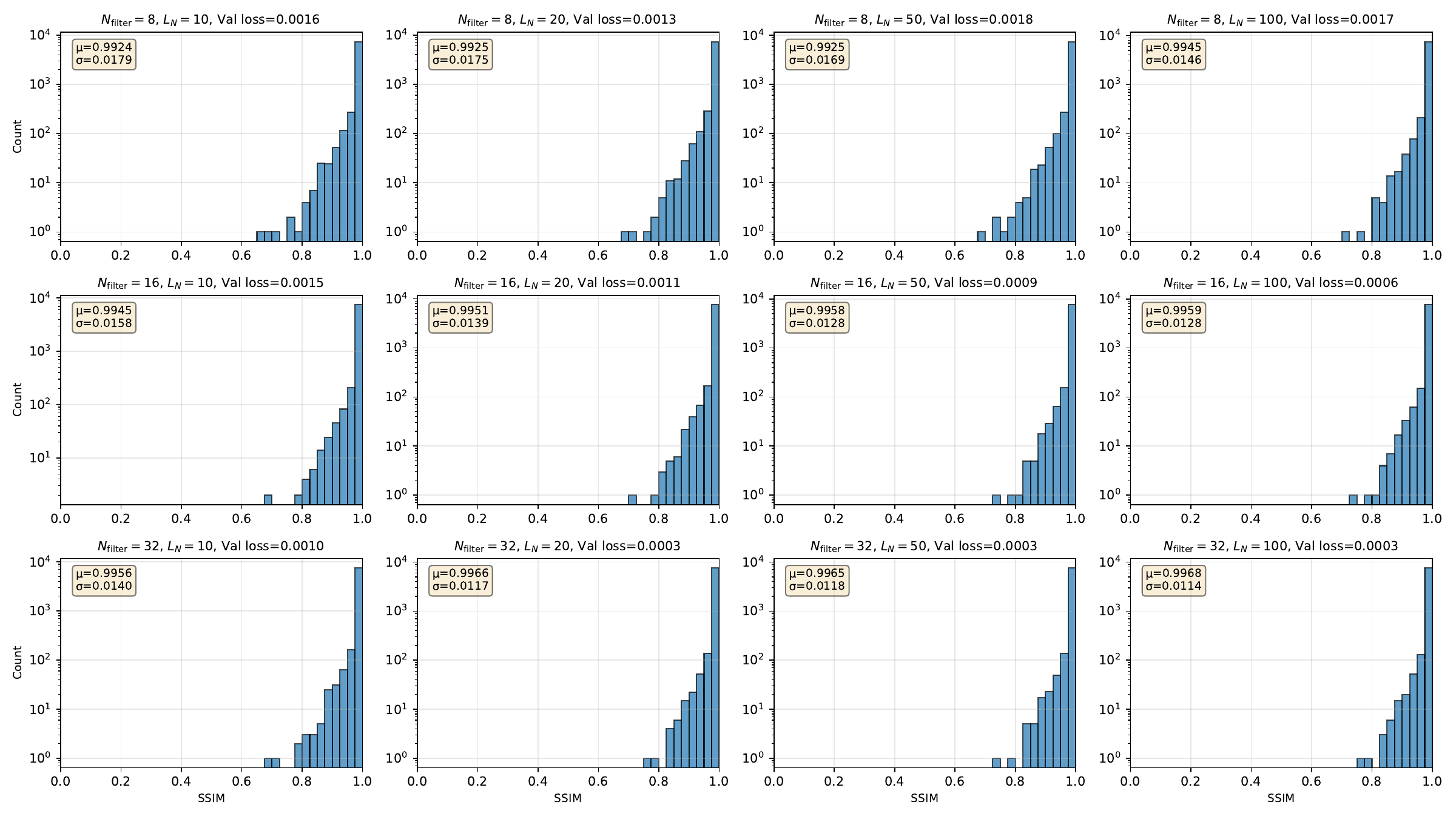}
    \caption{Histograms for the SSIM scores and validation losses for a grid of autoencoder configurations. $N_\mathrm{filter}$ refers to the number of filters per convolutional layer and $L_N$ is the size of the latent space.}
    \label{fig:ssim_grid}
\end{figure*}

We investigated the network configuration by running a grid of models, varying the $N_{\rm filter}$ and $L_N$, comparing both the validation and the SSIM distribution (see Figure~\ref{fig:ssim_grid}). It is apparent that even the model with $N_\mathrm{filter}=8$ and $L_N=10$ produces a relatively good distribution of SSIM scores. The distribution narrows significantly as the size of the latent space increases, and then as the number of filters increases. The validation loss drops significantly between the $N_\mathrm{filter}=32$ and $L_N=10$ and the $N_\mathrm{filter}=32$ and $L_N=20$ model, but then plateaus, with only vary marginal improvements in the SSIM distribution. We selected $N_\mathrm{filter}=32$ and $L_N=100$ as our final configuration.

\subsection{The latent space decoder}

Once the autoencoder has been trained, a fully-connected feed-forward neural network is used to learn the relationship between the disc input parameters and the latent space vectors (see Figure~\ref{fig:decoder}). This is a similar problem to the SED generator, with the same number of inputs and the same number of outputs. We therefore select a network configuration with two hidden layers and investigate the number of neurons per layer. Once again we use the validation loss as our principle figure of merit, but we also look at the SSIM distribution as a secondary indicator as to the quality of the image reproduction. 

\begin{figure*}
    \includegraphics[width=180mm]{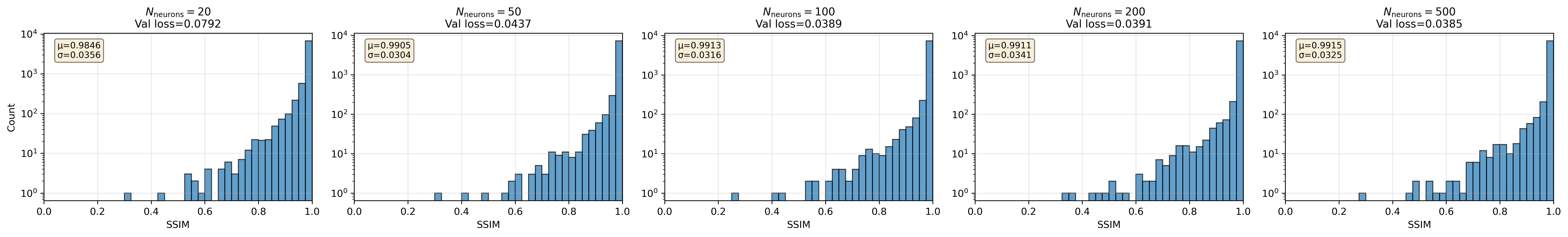}
    \caption{Histograms for the SSIM scores and validation losses for a grid of latent space/decoder configurations. $N_\mathrm{neurons}$ refers to the number of neurons per hidden layer.}
    \label{fig:ls_grid_row}
\end{figure*}

We searched 20, 50, 100, 200, and 500 neurons per hidden layer, and we plot the resulting SSIM distributions in Figure~\ref{fig:ls_grid_row} along with the resulting validation losses. We find that above 50 neurons per layer the validation loss begins to plateau, although the mean of the SSIM distribution increases at 500 neurons and the validation loss is smaller. We note that the training loss is significantly less than the validation loss for models with greater than 50 neurons per layer, although the validation loss is not increasing with epoch as might be expected for over-fitting. We therefore selected the 500 neuron/layer configuration. We conclude that any improvement in the reproduction of the images would need a larger training set, possibly with reduced noise.

\section{Summary of results}
\label{appendix_b}

\begin{figure*}
    \includegraphics[width=180mm]{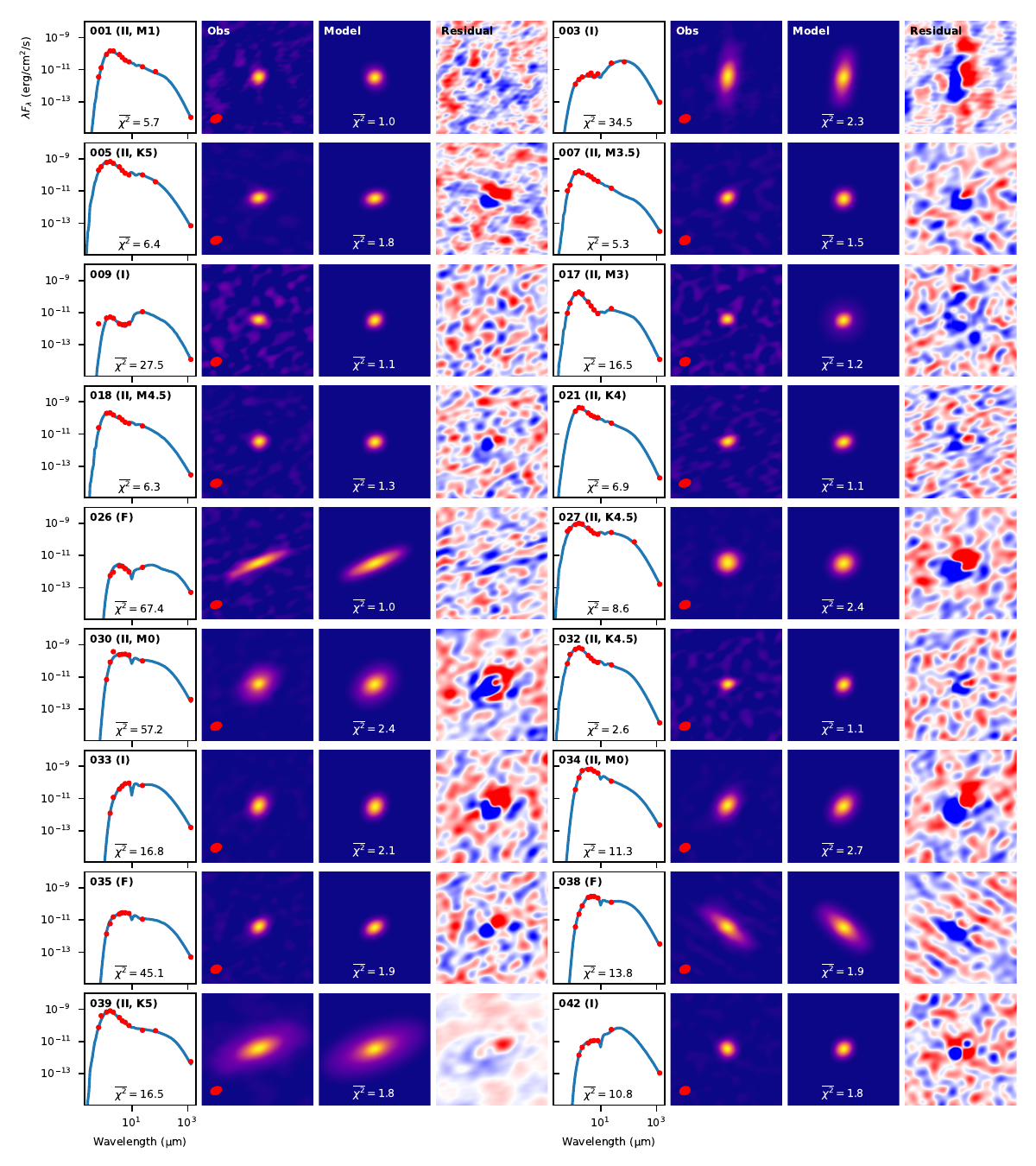}
	
    \caption{The results of the simultaneous fits to the images and SEDs. There are two columns of four panels, with each set of four panels corresponding to an individual object. The first panel shows the observed SED (red points) along with the best fit (blue line). The object number, its spectral class, and spectral type if available, are included in the plot along with the reduced chi-squared value of the SED fit. The second panel shows the 1.3\,mm ALMA image along with the beam (red ellipse). The third panel shows the model image, and is labelled with the reduced chisq-squared value of the image fit. The images are normalised and the colour scale goes from 0 (minimum) to 1 (maximum). The final panel shows the residual images (observation minus model) scaled between $-3\sigma$ (blue) to $+3\sigma$ (red).}
    \label{fig:full_plot}
\end{figure*}
\begin{figure*}
    \includegraphics[width=180mm]{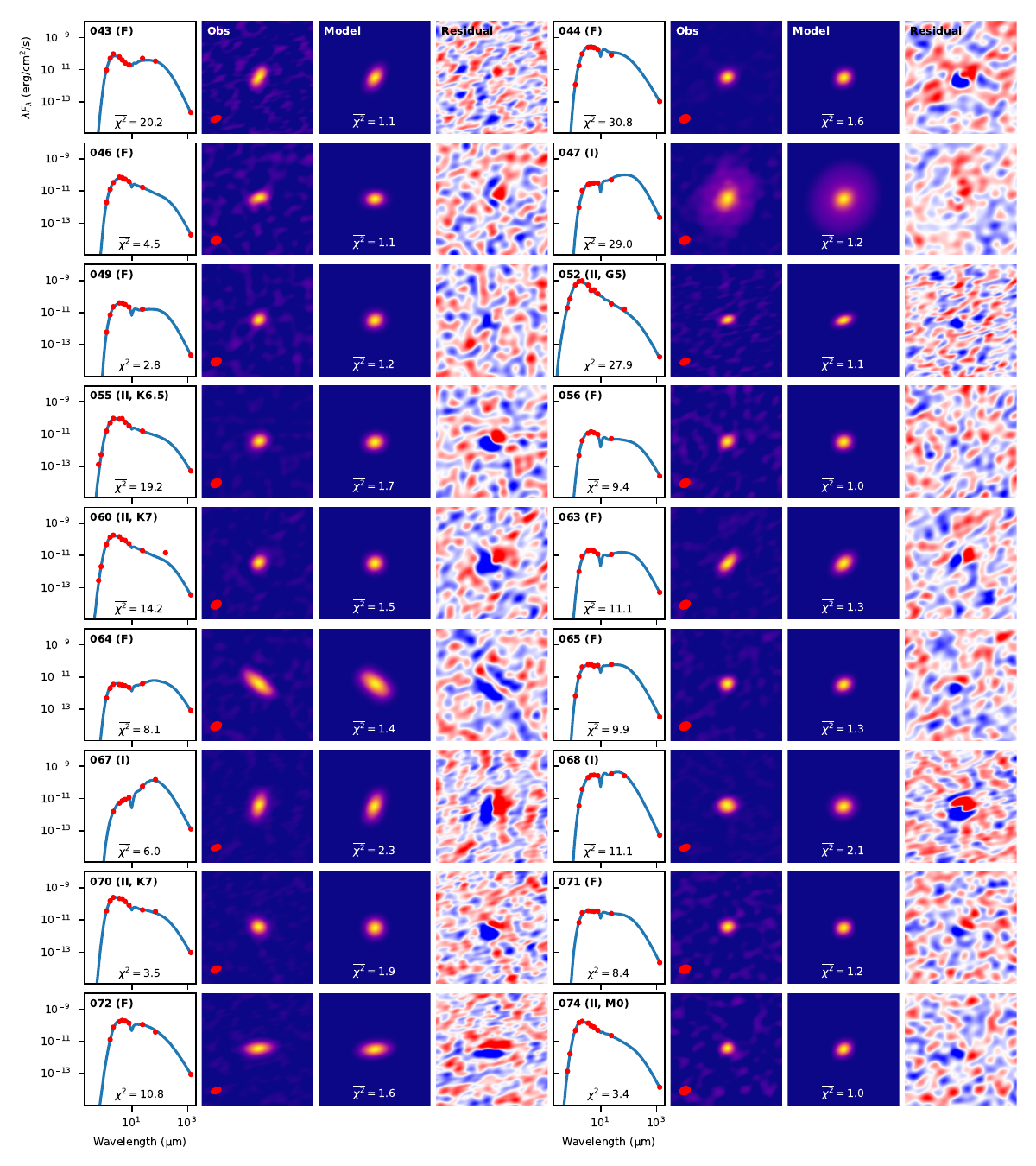}
	\addtocounter{figure}{-1}
    \caption{Continued.}
    \label{fig:full_plot}
\end{figure*}
\begin{figure*}
    \includegraphics[width=180mm]{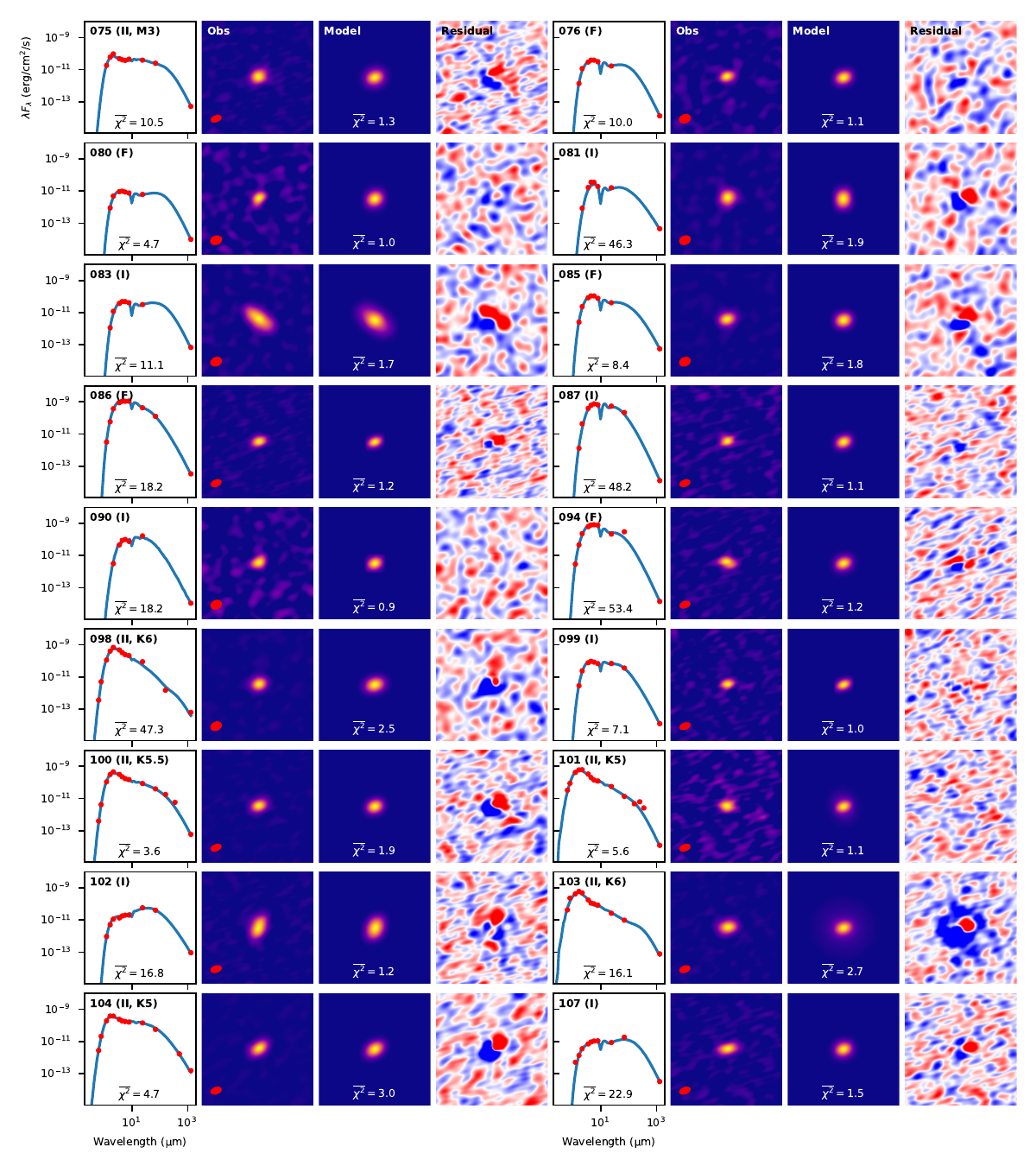}
	\addtocounter{figure}{-1}
    \caption{Continued.}
    \label{fig:full_plot}
\end{figure*}
\begin{figure*}
    \includegraphics[width=180mm]{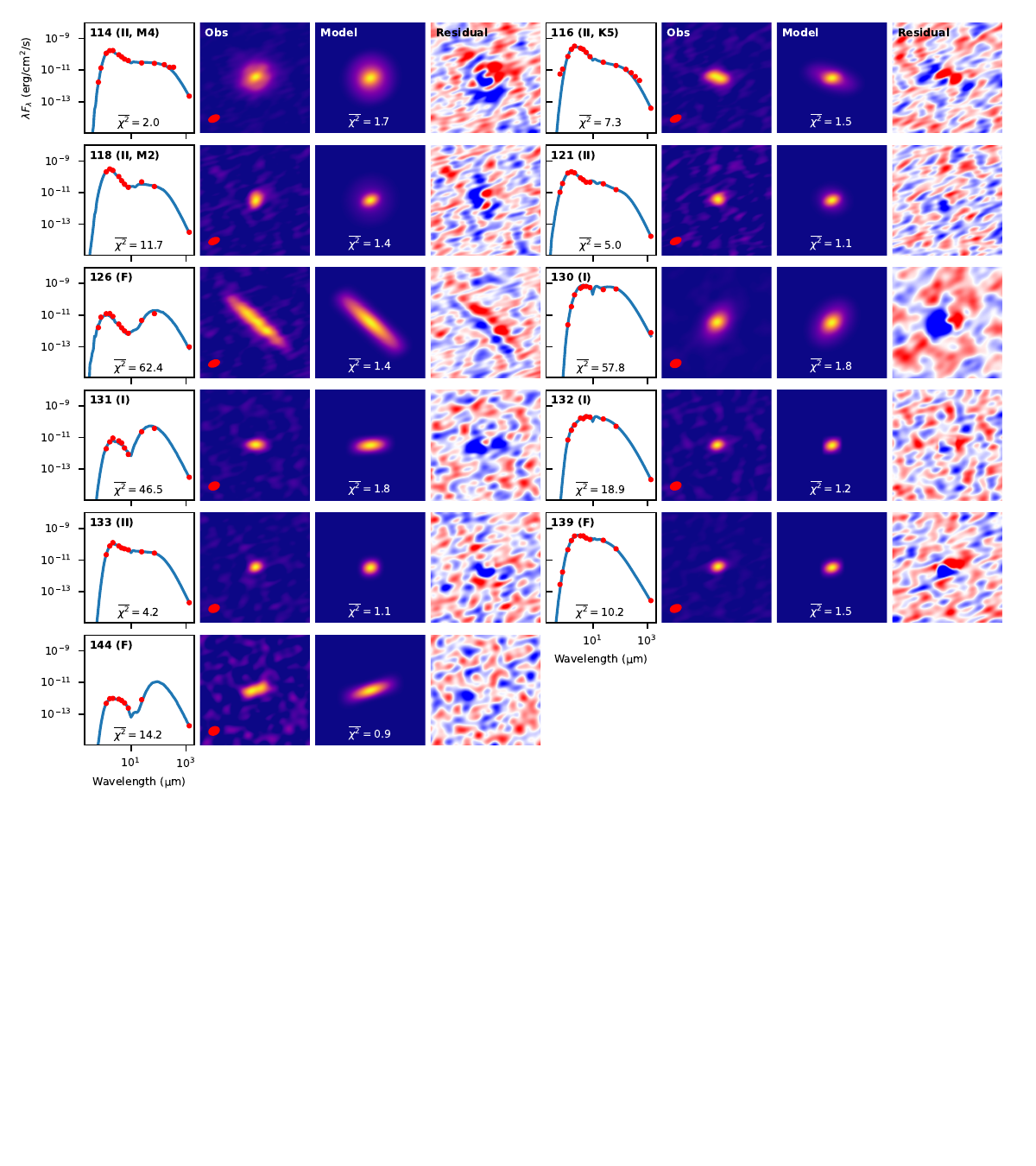}
	\addtocounter{figure}{-1}
    \caption{Concluded.}
    \label{fig:full_plot}
\end{figure*}

\begin{landscape}
\begin{table}
\caption{A summary of the results of the simultaneous fits to the SEDs and images. The uncertainties are derived from the MCMC posterior distributions. The first column gives the ODISEA object ID. The spectral types are taken from the sources described in \protect\cite{Cieza2019} supplemented by \protect\cite{Esplin2018} (IDs 1, 7, 11, 14, 20, 23, 25) and \protect\cite{Rydgren1980} (ID 140).}
\label{tab:full_results}
\centering
\renewcommand{\arraystretch}{1.4}
\begin{tabular}{lllllllllllllllll}
\hline
ID & Spec. & Spec. & $T_\mathrm{eff}$ & $A_V$ & $R_*$ & $M_\mathrm{dust}$ & $r_\mathrm{in}$ & $r_\mathrm{out}$ & $h/r$ & $\gamma$ & $\beta$ & $\xi$ & $i$ & $\mathrm{PA}$ & $\chi^2_\mathrm{SED}$ & $\chi^2_\mathrm{img}$ \\
 & class & type & (K) & (mag) & ($R_\odot$) & ($M_\oplus$) & ($R_\odot$) & (au) & & & & & ($^\circ$) & ($^\circ$) & & \\
\hline
1 & II & M1 & $3630$ & $\phantom{0}6.5^{+\phantom{0}0.5}_{-\phantom{0}0.4}$ & $1.6^{+0.1}_{-0.1}$ & $\phantom{00}0.9^{+\phantom{00}0.2}_{-\phantom{00}0.2}$ & $23^{+15}_{-9}$ & $\phantom{0}30^{+20}_{-13}$ & $0.07^{+0.02}_{-0.01}$ & $1.18^{+0.22}_{-0.35}$ & $1.20^{+0.06}_{-0.05}$ & $0.11^{+0.06}_{-0.05}$ & $\phantom{0}61^{+42}_{-33}$ & $\phantom{0}21^{+\phantom{0}19}_{-\phantom{0}53}$ & $6$ & $1.0$ \\
3 & I &  & $4970^{+300}_{-300}$ & $\phantom{0}8.9^{+\phantom{0}0.9}_{-\phantom{0}0.9}$ & $0.9^{+0.1}_{-0.1}$ & $\phantom{0}12\phantom{.0}^{+\phantom{00}3\phantom{.0}}_{-\phantom{00}2\phantom{.0}}$ & $24^{+67}_{-10}$ & $\phantom{0}68^{+\phantom{0}9}_{-\phantom{0}9}$ & $0.18^{+0.01}_{-0.02}$ & $1.08^{+0.18}_{-0.22}$ & $1.38^{+0.10}_{-0.12}$ & $0.04^{+0.01}_{-0.02}$ & $\phantom{0}76^{+\phantom{0}1}_{-\phantom{0}2}$ & $174^{+\phantom{00}3}_{-\phantom{00}3}$ & $35$ & $2.3$ \\
5 & II & K5    & $4140$ & $\phantom{0}2.5^{+\phantom{0}0.6}_{-\phantom{0}0.5}$ & $2.0^{+0.2}_{-0.1}$ & $\phantom{0}94\phantom{.0}^{+100\phantom{.0}}_{-\phantom{0}70\phantom{.0}}$ & $19^{+13}_{-6}$ & $\phantom{0}19^{+\phantom{0}3}_{-\phantom{0}3}$ & $0.09^{+0.02}_{-0.02}$ & $0.97^{+0.32}_{-0.30}$ & $1.15^{+0.06}_{-0.03}$ & $0.05^{+0.07}_{-0.03}$ & $\phantom{0}58^{+\phantom{0}7}_{-15}$ & $\phantom{0}90^{+\phantom{0}19}_{-\phantom{0}22}$ & $6$ & $1.8$ \\
7 & II & M3.5  & $3260$ & $\phantom{0}4.4^{+\phantom{0}0.5}_{-\phantom{0}0.4}$ & $1.7^{+0.1}_{-0.1}$ & $\phantom{0}13\phantom{.0}^{+\phantom{0}50\phantom{.0}}_{-\phantom{00}7\phantom{.0}}$ & $20^{+9}_{-6}$ & $\phantom{0}18^{+\phantom{0}5}_{-\phantom{0}3}$ & $0.06^{+0.01}_{-0.01}$ & $0.91^{+0.35}_{-0.27}$ & $1.07^{+0.04}_{-0.03}$ & $0.05^{+0.05}_{-0.03}$ & $\phantom{0}44^{+15}_{-18}$ & $\phantom{0}14^{+\phantom{0}74}_{-\phantom{0}69}$ & $5$ & $1.5$ \\
9 & I &  & $3080^{+100}_{-60}$ & $\phantom{0}8.7^{+\phantom{0}0.6}_{-\phantom{0}0.6}$ & $0.5^{+0.1}_{-0.1}$ & $\phantom{00}1.1^{+\phantom{00}0.5}_{-\phantom{00}0.3}$ & $88^{+13}_{-13}$ & $\phantom{0}14^{+12}_{-\phantom{0}7}$ & $0.16^{+0.03}_{-0.04}$ & $0.69^{+0.48}_{-0.15}$ & $1.07^{+0.05}_{-0.05}$ & $0.00^{+0.01}_{-0.01}$ & $\phantom{00}1^{+\phantom{0}3}_{-\phantom{0}1}$ & $\phantom{00}4^{+\phantom{0}59}_{-\phantom{0}22}$ & $27$ & $1.1$ \\
17 & II & M3    & $3360$ & $\phantom{0}3.6^{+\phantom{0}0.3}_{-\phantom{0}0.3}$ & $1.5^{+0.1}_{-0.1}$ & $\phantom{00}0.9^{+\phantom{00}0.2}_{-\phantom{00}0.2}$ & $97^{+8}_{-12}$ & $\phantom{0}51^{+\phantom{0}7}_{-17}$ & $0.07^{+0.02}_{-0.01}$ & $1.23^{+0.16}_{-0.25}$ & $1.44^{+0.05}_{-0.09}$ & $0.17^{+0.02}_{-0.06}$ & $177^{+\phantom{0}2}_{-\phantom{0}5}$ & $\phantom{0}14^{+\phantom{0}14}_{-\phantom{0}41}$ & $17$ & $1.2$ \\
18 & II & M4.5  & $3020$ & $\phantom{0}2.9^{+\phantom{0}0.8}_{-\phantom{0}0.6}$ & $2.0^{+0.1}_{-0.1}$ & $\phantom{0}24\phantom{.0}^{+100\phantom{.0}}_{-\phantom{0}20\phantom{.0}}$ & $22^{+13}_{-8}$ & $\phantom{0}13^{+\phantom{0}4}_{-\phantom{0}3}$ & $0.09^{+0.02}_{-0.02}$ & $1.13^{+0.28}_{-0.41}$ & $1.19^{+0.07}_{-0.06}$ & $0.08^{+0.08}_{-0.06}$ & $\phantom{0}49^{+17}_{-25}$ & $\phantom{0}19^{+\phantom{0}20}_{-\phantom{0}90}$ & $6$ & $1.3$ \\
21 & II & K4    & $4330$ & $\phantom{0}6.6^{+\phantom{0}0.7}_{-\phantom{0}0.8}$ & $2.1^{+0.1}_{-0.1}$ & $\phantom{00}1.1^{+\phantom{00}0.3}_{-\phantom{00}0.2}$ & $22^{+27}_{-9}$ & $\phantom{0}22^{+24}_{-11}$ & $0.06^{+0.02}_{-0.01}$ & $1.27^{+0.16}_{-0.31}$ & $1.15^{+0.05}_{-0.04}$ & $0.14^{+0.04}_{-0.06}$ & $\phantom{0}10^{+21}_{-\phantom{0}8}$ & $\phantom{00}5^{+\phantom{0}65}_{-\phantom{0}72}$ & $7$ & $1.1$ \\
26 & F &  & $3380^{+300}_{-300}$ & $23.5^{+\phantom{0}6.0}_{-\phantom{0}8.0}$ & $0.6^{+0.1}_{-0.1}$ & $\phantom{0}43\phantom{.0}^{+\phantom{0}50\phantom{.0}}_{-\phantom{0}20\phantom{.0}}$ & $91^{+13}_{-25}$ & $\phantom{0}78^{+13}_{-13}$ & $0.10^{+0.04}_{-0.03}$ & $0.89^{+0.38}_{-0.28}$ & $1.42^{+0.06}_{-0.09}$ & $0.14^{+0.04}_{-0.04}$ & $\phantom{0}80^{+\phantom{0}2}_{-\phantom{0}2}$ & $114^{+\phantom{00}3}_{-\phantom{00}4}$ & $67$ & $1.0$ \\
27 & II & K4.5  & $4240$ & $\phantom{0}2.9^{+\phantom{0}0.9}_{-\phantom{0}0.6}$ & $2.4^{+0.3}_{-0.2}$ & $\phantom{0}11\phantom{.0}^{+\phantom{00}4\phantom{.0}}_{-\phantom{00}2\phantom{.0}}$ & $23^{+21}_{-9}$ & $\phantom{0}31^{+\phantom{0}4}_{-\phantom{0}3}$ & $0.14^{+0.03}_{-0.03}$ & $0.68^{+0.32}_{-0.14}$ & $1.22^{+0.06}_{-0.04}$ & $0.06^{+0.07}_{-0.04}$ & $152^{+14}_{-10}$ & $134^{+133}_{-114}$ & $9$ & $2.4$ \\
30 & II & M0    & $3770$ & $22.2^{+\phantom{0}0.4}_{-\phantom{0}0.5}$ & $2.9^{+0.1}_{-0.1}$ & $\phantom{0}50\phantom{.0}^{+\phantom{0}20\phantom{.0}}_{-\phantom{0}10\phantom{.0}}$ & $18^{+25}_{-6}$ & $\phantom{0}56^{+\phantom{0}5}_{-\phantom{0}5}$ & $0.15^{+0.02}_{-0.02}$ & $1.25^{+0.18}_{-0.19}$ & $1.10^{+0.03}_{-0.02}$ & $0.01^{+0.02}_{-0.01}$ & $\phantom{0}45^{+\phantom{0}4}_{-\phantom{0}5}$ & $128^{+\phantom{00}7}_{-\phantom{00}7}$ & $57$ & $2.4$ \\
32 & II & K4.5  & $4240$ & $\phantom{0}3.8^{+\phantom{0}0.4}_{-\phantom{0}0.3}$ & $2.2^{+0.1}_{-0.1}$ & $\phantom{00}0.8^{+\phantom{00}0.2}_{-\phantom{00}0.2}$ & $20^{+20}_{-7}$ & $\phantom{0}26^{+24}_{-12}$ & $0.07^{+0.02}_{-0.01}$ & $1.22^{+0.20}_{-0.43}$ & $1.30^{+0.08}_{-0.07}$ & $0.13^{+0.05}_{-0.06}$ & $\phantom{0}56^{+18}_{-31}$ & $173^{+\phantom{0}83}_{-\phantom{0}60}$ & $3$ & $1.1$ \\
33 & I &  & $3950^{+400}_{-400}$ & $42.5^{+\phantom{0}1.0}_{-\phantom{0}1.0}$ & $2.0^{+0.4}_{-0.3}$ & $\phantom{0}90\phantom{.0}^{+100\phantom{.0}}_{-\phantom{0}60\phantom{.0}}$ & $97^{+8}_{-13}$ & $\phantom{0}26^{+\phantom{0}2}_{-\phantom{0}3}$ & $0.17^{+0.02}_{-0.03}$ & $1.04^{+0.29}_{-0.34}$ & $1.24^{+0.05}_{-0.06}$ & $0.13^{+0.05}_{-0.05}$ & $136^{+15}_{-\phantom{0}7}$ & $161^{+\phantom{0}14}_{-\phantom{0}12}$ & $17$ & $2.1$ \\
34 & II & M0    & $3770$ & $16.3^{+\phantom{0}0.6}_{-\phantom{0}0.7}$ & $2.9^{+0.1}_{-0.1}$ & $\phantom{0}28\phantom{.0}^{+\phantom{00}8\phantom{.0}}_{-\phantom{00}6\phantom{.0}}$ & $20^{+6}_{-5}$ & $\phantom{0}40^{+\phantom{0}4}_{-\phantom{0}3}$ & $0.17^{+0.01}_{-0.01}$ & $0.82^{+0.21}_{-0.20}$ & $1.02^{+0.02}_{-0.01}$ & $0.02^{+0.02}_{-0.01}$ & $\phantom{0}52^{+\phantom{0}3}_{-\phantom{0}3}$ & $145^{+\phantom{00}7}_{-\phantom{00}6}$ & $11$ & $2.7$ \\
35 & F &  & $3160^{+400}_{-100}$ & $19.4^{+\phantom{0}2.0}_{-\phantom{0}2.0}$ & $1.3^{+0.2}_{-0.3}$ & $\phantom{0}12\phantom{.0}^{+\phantom{0}40\phantom{.0}}_{-\phantom{00}5\phantom{.0}}$ & $24^{+11}_{-7}$ & $\phantom{0}24^{+\phantom{0}5}_{-\phantom{0}4}$ & $0.14^{+0.03}_{-0.03}$ & $0.71^{+0.32}_{-0.16}$ & $1.13^{+0.04}_{-0.03}$ & $0.01^{+0.03}_{-0.01}$ & $\phantom{0}62^{+\phantom{0}5}_{-\phantom{0}5}$ & $122^{+\phantom{0}17}_{-\phantom{0}23}$ & $45$ & $1.9$ \\
38 & F &  & $6510^{+500}_{-600}$ & $25.3^{+\phantom{0}1.0}_{-\phantom{0}2.0}$ & $1.5^{+0.2}_{-0.2}$ & $\phantom{0}30\phantom{.0}^{+\phantom{00}7\phantom{.0}}_{-\phantom{00}5\phantom{.0}}$ & $85^{+15}_{-19}$ & $\phantom{0}76^{+\phantom{0}7}_{-\phantom{0}6}$ & $0.11^{+0.03}_{-0.02}$ & $1.13^{+0.21}_{-0.22}$ & $1.20^{+0.03}_{-0.03}$ & $0.03^{+0.03}_{-0.02}$ & $\phantom{0}70^{+\phantom{0}2}_{-\phantom{0}2}$ & $\phantom{0}49^{+\phantom{00}2}_{-\phantom{00}2}$ & $14$ & $1.9$ \\
39 & II & K5    & $4140$ & $\phantom{0}4.1^{+\phantom{0}0.4}_{-\phantom{0}0.4}$ & $2.5^{+0.1}_{-0.2}$ & $130\phantom{.0}^{+\phantom{0}30\phantom{.0}}_{-\phantom{0}30\phantom{.0}}$ & $45^{+17}_{-23}$ & $125^{+12}_{-\phantom{0}9}$ & $0.06^{+0.01}_{-0.01}$ & $1.34^{+0.12}_{-0.37}$ & $1.14^{+0.02}_{-0.02}$ & $0.00^{+0.01}_{-0.01}$ & $\phantom{0}67^{+\phantom{0}2}_{-\phantom{0}2}$ & $110^{+\phantom{00}2}_{-\phantom{00}2}$ & $16$ & $1.8$ \\
42 & I &  & $5140^{+300}_{-500}$ & $32.4^{+\phantom{0}2.0}_{-\phantom{0}1.0}$ & $0.6^{+0.2}_{-0.1}$ & $\phantom{0}77\phantom{.0}^{+100\phantom{.0}}_{-\phantom{0}50\phantom{.0}}$ & $85^{+15}_{-30}$ & $\phantom{0}15^{+\phantom{0}2}_{-\phantom{0}2}$ & $0.17^{+0.02}_{-0.02}$ & $0.98^{+0.28}_{-0.29}$ & $1.45^{+0.04}_{-0.04}$ & $0.02^{+0.01}_{-0.01}$ & $\phantom{00}3^{+\phantom{0}5}_{-\phantom{0}2}$ & $\phantom{0}70^{+\phantom{0}68}_{-\phantom{0}92}$ & $11$ & $1.8$ \\
43 & F &  & $4260^{+400}_{-300}$ & $17.0^{+\phantom{0}0.8}_{-\phantom{0}0.8}$ & $1.5^{+0.1}_{-0.1}$ & $\phantom{00}1.3^{+\phantom{00}0.2}_{-\phantom{00}0.2}$ & $23^{+14}_{-8}$ & $\phantom{0}32^{+\phantom{0}4}_{-\phantom{0}4}$ & $0.12^{+0.04}_{-0.04}$ & $0.58^{+0.15}_{-0.06}$ & $1.49^{+0.01}_{-0.02}$ & $0.11^{+0.04}_{-0.04}$ & $116^{+12}_{-\phantom{0}8}$ & $159^{+\phantom{00}9}_{-\phantom{0}10}$ & $20$ & $1.1$ \\
44 & F &  & $5600^{+1000}_{-1000}$ & $29.5^{+\phantom{0}2.0}_{-\phantom{0}2.0}$ & $1.7^{+0.5}_{-0.7}$ & $\phantom{0}22\phantom{.0}^{+\phantom{0}70\phantom{.0}}_{-\phantom{0}10\phantom{.0}}$ & $16^{+9}_{-4}$ & $\phantom{0}15^{+\phantom{0}2}_{-\phantom{0}2}$ & $0.12^{+0.05}_{-0.05}$ & $0.70^{+0.28}_{-0.15}$ & $1.08^{+0.04}_{-0.04}$ & $0.02^{+0.05}_{-0.02}$ & $147^{+16}_{-13}$ & $153^{+116}_{-103}$ & $31$ & $1.6$ \\
46 & F &  & $4100^{+800}_{-600}$ & $22.1^{+\phantom{0}2.0}_{-\phantom{0}2.0}$ & $1.3^{+0.4}_{-0.4}$ & $\phantom{00}2.6^{+\phantom{00}3\phantom{.0}}_{-\phantom{00}0.9}$ & $19^{+11}_{-6}$ & $\phantom{0}28^{+19}_{-12}$ & $0.14^{+0.04}_{-0.05}$ & $1.05^{+0.30}_{-0.34}$ & $1.11^{+0.06}_{-0.05}$ & $0.14^{+0.04}_{-0.07}$ & $117^{+14}_{-\phantom{0}8}$ & $\phantom{0}83^{+173}_{-\phantom{0}87}$ & $4$ & $1.1$ \\
47 & I &  & $5990^{+1000}_{-1000}$ & $44.0^{+\phantom{0}1.0}_{-\phantom{0}2.0}$ & $1.3^{+0.4}_{-0.4}$ & $\phantom{0}19\phantom{.0}^{+\phantom{00}3\phantom{.0}}_{-\phantom{00}3\phantom{.0}}$ & $78^{+22}_{-46}$ & $\phantom{0}84^{+12}_{-\phantom{0}9}$ & $0.11^{+0.06}_{-0.04}$ & $0.64^{+0.14}_{-0.10}$ & $1.41^{+0.06}_{-0.06}$ & $0.15^{+0.03}_{-0.07}$ & $151^{+18}_{-15}$ & $151^{+151}_{-\phantom{0}25}$ & $29$ & $1.2$ \\
49 & F &  & $3310^{+400}_{-200}$ & $28.5^{+\phantom{0}1.0}_{-\phantom{0}1.0}$ & $2.2^{+0.3}_{-0.4}$ & $\phantom{00}1.4^{+\phantom{00}0.4}_{-\phantom{00}0.3}$ & $46^{+36}_{-26}$ & $\phantom{0}23^{+24}_{-10}$ & $0.08^{+0.03}_{-0.02}$ & $1.17^{+0.23}_{-0.36}$ & $1.33^{+0.07}_{-0.07}$ & $0.11^{+0.07}_{-0.07}$ & $\phantom{0}31^{+23}_{-20}$ & $\phantom{0}76^{+\phantom{0}26}_{-114}$ & $3$ & $1.2$ \\
52 & II & G5    & $5500$ & $\phantom{0}7.0^{+\phantom{0}0.4}_{-\phantom{0}0.4}$ & $2.1^{+0.2}_{-0.1}$ & $\phantom{00}1.7^{+\phantom{00}2\phantom{.0}}_{-\phantom{00}0.5}$ & $13^{+4}_{-1}$ & $\phantom{0}19^{+\phantom{0}4}_{-\phantom{0}4}$ & $0.05^{+0.01}_{-0.01}$ & $0.74^{+0.41}_{-0.19}$ & $1.05^{+0.02}_{-0.02}$ & $0.07^{+0.04}_{-0.03}$ & $107^{+\phantom{0}5}_{-\phantom{0}3}$ & $102^{+\phantom{0}11}_{-\phantom{0}13}$ & $28$ & $1.1$ \\
55 & II & K6.5  & $4000$ & $11.8^{+\phantom{0}1.0}_{-\phantom{0}0.9}$ & $1.0^{+0.2}_{-0.1}$ & $\phantom{0}18\phantom{.0}^{+\phantom{0}40\phantom{.0}}_{-\phantom{00}9\phantom{.0}}$ & $13^{+2}_{-1}$ & $\phantom{0}21^{+\phantom{0}5}_{-\phantom{0}3}$ & $0.11^{+0.02}_{-0.02}$ & $0.72^{+0.21}_{-0.15}$ & $1.05^{+0.02}_{-0.03}$ & $0.02^{+0.02}_{-0.02}$ & $132^{+11}_{-10}$ & $\phantom{0}84^{+\phantom{0}11}_{-109}$ & $19$ & $1.7$ \\
56 & F &  & $4510^{+400}_{-400}$ & $39.4^{+\phantom{0}2.0}_{-\phantom{0}2.0}$ & $0.7^{+0.2}_{-0.1}$ & $\phantom{00}6.1^{+\phantom{0}30\phantom{.0}}_{-\phantom{00}3\phantom{.0}}$ & $14^{+6}_{-3}$ & $\phantom{0}17^{+\phantom{0}9}_{-\phantom{0}5}$ & $0.09^{+0.04}_{-0.03}$ & $1.17^{+0.24}_{-0.40}$ & $1.17^{+0.05}_{-0.05}$ & $0.10^{+0.07}_{-0.06}$ & $\phantom{0}39^{+21}_{-22}$ & $\phantom{0}63^{+\phantom{0}23}_{-\phantom{0}98}$ & $9$ & $1.0$ \\
60 & II & K7    & $3970$ & $10.7^{+\phantom{0}0.5}_{-\phantom{0}0.5}$ & $1.7^{+0.1}_{-0.1}$ & $\phantom{00}6.3^{+\phantom{0}10\phantom{.0}}_{-\phantom{00}2\phantom{.0}}$ & $16^{+7}_{-4}$ & $\phantom{0}20^{+\phantom{0}9}_{-\phantom{0}5}$ & $0.06^{+0.01}_{-0.01}$ & $0.91^{+0.33}_{-0.28}$ & $1.06^{+0.03}_{-0.03}$ & $0.04^{+0.05}_{-0.03}$ & $142^{+18}_{-15}$ & $\phantom{0}44^{+\phantom{0}53}_{-\phantom{0}83}$ & $14$ & $1.5$ \\
63 & F &  & $3670^{+400}_{-400}$ & $41.0^{+\phantom{0}1.0}_{-\phantom{0}1.0}$ & $1.8^{+0.3}_{-0.3}$ & $\phantom{00}4.9^{+\phantom{0}10\phantom{.0}}_{-\phantom{00}2\phantom{.0}}$ & $61^{+35}_{-38}$ & $\phantom{0}24^{+\phantom{0}9}_{-\phantom{0}6}$ & $0.06^{+0.02}_{-0.01}$ & $0.87^{+0.41}_{-0.29}$ & $1.38^{+0.06}_{-0.06}$ & $0.07^{+0.08}_{-0.05}$ & $124^{+27}_{-14}$ & $140^{+\phantom{0}92}_{-175}$ & $11$ & $1.3$ \\
64 & F &  & $3480^{+300}_{-300}$ & $15.9^{+\phantom{0}2.0}_{-\phantom{0}3.0}$ & $0.6^{+0.1}_{-0.1}$ & $\phantom{0}83\phantom{.0}^{+100\phantom{.0}}_{-\phantom{0}50\phantom{.0}}$ & $52^{+24}_{-19}$ & $\phantom{0}43^{+\phantom{0}5}_{-\phantom{0}4}$ & $0.12^{+0.03}_{-0.02}$ & $0.92^{+0.35}_{-0.30}$ & $1.38^{+0.06}_{-0.06}$ & $0.03^{+0.03}_{-0.02}$ & $\phantom{0}71^{+\phantom{0}3}_{-\phantom{0}3}$ & $\phantom{0}49^{+\phantom{00}7}_{-\phantom{00}6}$ & $8$ & $1.4$ \\
65 & F &  & $4000^{+500}_{-400}$ & $29.7^{+\phantom{0}1.0}_{-\phantom{0}0.9}$ & $2.2^{+0.4}_{-0.3}$ & $\phantom{00}1.4^{+\phantom{00}0.5}_{-\phantom{00}0.3}$ & $68^{+29}_{-38}$ & $\phantom{0}12^{+\phantom{0}9}_{-\phantom{0}5}$ & $0.10^{+0.04}_{-0.03}$ & $1.09^{+0.30}_{-0.39}$ & $1.43^{+0.05}_{-0.06}$ & $0.15^{+0.04}_{-0.07}$ & $\phantom{00}1^{+\phantom{0}1}_{-\phantom{0}1}$ & $\phantom{0}14^{+\phantom{0}64}_{-\phantom{0}67}$ & $10$ & $1.3$ \\
\hline
\end{tabular}
\renewcommand{\arraystretch}{1.0}
\end{table}
\end{landscape}

\begin{landscape}
\begin{table}
\centering
\renewcommand{\arraystretch}{1.4}
\begin{tabular}{lllllllllllllllll}
\hline
ID & Spec. & Spec. & $T_\mathrm{eff}$ & $A_V$ & $R_*$ & $M_\mathrm{dust}$ & $r_\mathrm{in}$ & $r_\mathrm{out}$ & $h/r$ & $\gamma$ & $\beta$ & $\xi$ & $i$ & $\mathrm{PA}$ & $\chi^2_\mathrm{SED}$ & $\chi^2_\mathrm{img}$ \\
 & class & type & (K) & (mag) & ($R_\odot$) & ($M_\oplus$) & ($R_\odot$) & (au) & & & & & ($^\circ$) & ($^\circ$) & & \\
\hline
67 & I &  & $5400^{+1000}_{-800}$ & $29.0^{+\phantom{0}4.0}_{-\phantom{0}5.0}$ & $1.9^{+0.7}_{-0.6}$ & $\phantom{00}9.8^{+\phantom{00}2\phantom{.0}}_{-\phantom{00}2\phantom{.0}}$ & $38^{+36}_{-17}$ & $\phantom{0}39^{+\phantom{0}6}_{-\phantom{0}5}$ & $0.18^{+0.01}_{-0.02}$ & $0.90^{+0.34}_{-0.27}$ & $1.27^{+0.09}_{-0.07}$ & $0.00^{+0.01}_{-0.01}$ & $\phantom{0}73^{+\phantom{0}2}_{-\phantom{0}2}$ & $165^{+\phantom{00}6}_{-\phantom{00}6}$ & $6$ & $2.3$ \\
68 & I &  & $7040^{+600}_{-800}$ & $46.6^{+\phantom{0}1.0}_{-\phantom{0}1.0}$ & $2.7^{+0.2}_{-0.4}$ & $\phantom{00}1.3^{+\phantom{00}0.2}_{-\phantom{00}0.2}$ & $44^{+37}_{-24}$ & $\phantom{0}27^{+\phantom{0}4}_{-\phantom{0}3}$ & $0.09^{+0.05}_{-0.03}$ & $0.62^{+0.16}_{-0.09}$ & $1.35^{+0.07}_{-0.06}$ & $0.09^{+0.06}_{-0.05}$ & $152^{+17}_{-15}$ & $\phantom{0}52^{+\phantom{0}16}_{-\phantom{0}39}$ & $11$ & $2.1$ \\
70 & II & K7    & $3970$ & $15.5^{+\phantom{0}0.7}_{-\phantom{0}0.8}$ & $2.6^{+0.2}_{-0.2}$ & $\phantom{0}14\phantom{.0}^{+\phantom{0}30\phantom{.0}}_{-\phantom{00}6\phantom{.0}}$ & $24^{+9}_{-8}$ & $\phantom{0}25^{+\phantom{0}7}_{-\phantom{0}4}$ & $0.06^{+0.01}_{-0.01}$ & $1.01^{+0.31}_{-0.33}$ & $1.10^{+0.03}_{-0.02}$ & $0.01^{+0.02}_{-0.01}$ & $138^{+23}_{-14}$ & $\phantom{0}22^{+\phantom{0}25}_{-\phantom{00}3}$ & $3$ & $1.9$ \\
71 & F &  & $3810^{+500}_{-500}$ & $28.3^{+\phantom{0}1.0}_{-\phantom{0}1.0}$ & $1.8^{+0.3}_{-0.3}$ & $\phantom{00}1.4^{+\phantom{00}2\phantom{.0}}_{-\phantom{00}0.4}$ & $89^{+13}_{-22}$ & $\phantom{0}13^{+\phantom{0}8}_{-\phantom{0}5}$ & $0.09^{+0.04}_{-0.02}$ & $0.80^{+0.43}_{-0.21}$ & $1.32^{+0.07}_{-0.07}$ & $0.15^{+0.04}_{-0.06}$ & $\phantom{0}45^{+21}_{-26}$ & $\phantom{0}43^{+\phantom{0}35}_{-\phantom{0}95}$ & $8$ & $1.2$ \\
72 & F &  & $7010^{+700}_{-800}$ & $11.5^{+\phantom{0}5.0}_{-\phantom{0}6.0}$ & $0.9^{+0.2}_{-0.1}$ & $\phantom{0}55\phantom{.0}^{+\phantom{0}90\phantom{.0}}_{-\phantom{0}30\phantom{.0}}$ & $17^{+7}_{-4}$ & $\phantom{0}38^{+\phantom{0}5}_{-\phantom{0}3}$ & $0.18^{+0.01}_{-0.03}$ & $0.61^{+0.19}_{-0.08}$ & $1.06^{+0.03}_{-0.03}$ & $0.16^{+0.03}_{-0.03}$ & $111^{+\phantom{0}3}_{-\phantom{0}3}$ & $\phantom{0}93^{+\phantom{00}6}_{-\phantom{00}6}$ & $11$ & $1.6$ \\
74 & II & M0    & $3770$ & $11.4^{+\phantom{0}0.4}_{-\phantom{0}0.4}$ & $2.0^{+0.1}_{-0.1}$ & $\phantom{00}1.3^{+\phantom{00}2\phantom{.0}}_{-\phantom{00}0.4}$ & $18^{+8}_{-5}$ & $\phantom{0}13^{+10}_{-\phantom{0}5}$ & $0.06^{+0.01}_{-0.01}$ & $1.13^{+0.26}_{-0.40}$ & $1.12^{+0.05}_{-0.04}$ & $0.10^{+0.06}_{-0.06}$ & $\phantom{0}53^{+18}_{-30}$ & $167^{+\phantom{0}80}_{-\phantom{0}69}$ & $3$ & $1.0$ \\
75 & II & M3    & $3360$ & $13.3^{+\phantom{0}0.6}_{-\phantom{0}0.6}$ & $1.9^{+0.1}_{-0.1}$ & $\phantom{00}4.0^{+\phantom{00}1\phantom{.0}}_{-\phantom{00}0.7}$ & $94^{+10}_{-15}$ & $\phantom{0}26^{+17}_{-\phantom{0}7}$ & $0.12^{+0.03}_{-0.03}$ & $1.13^{+0.26}_{-0.40}$ & $1.29^{+0.06}_{-0.06}$ & $0.14^{+0.04}_{-0.06}$ & $\phantom{0}13^{+17}_{-\phantom{0}9}$ & $\phantom{0}28^{+\phantom{0}51}_{-\phantom{0}68}$ & $11$ & $1.3$ \\
76 & F &  & $3670^{+400}_{-300}$ & $41.9^{+\phantom{0}1.0}_{-\phantom{0}2.0}$ & $2.2^{+0.4}_{-0.3}$ & $\phantom{00}0.6^{+\phantom{00}0.2}_{-\phantom{00}0.1}$ & $88^{+14}_{-26}$ & $\phantom{0}12^{+\phantom{0}9}_{-\phantom{0}5}$ & $0.07^{+0.03}_{-0.01}$ & $0.63^{+0.32}_{-0.10}$ & $1.22^{+0.07}_{-0.07}$ & $0.10^{+0.06}_{-0.06}$ & $150^{+22}_{-28}$ & $\phantom{0}86^{+\phantom{00}5}_{-103}$ & $10$ & $1.1$ \\
80 & F &  & $4110^{+700}_{-500}$ & $36.2^{+\phantom{0}1.0}_{-\phantom{0}1.0}$ & $1.0^{+0.2}_{-0.2}$ & $\phantom{00}0.7^{+\phantom{00}0.3}_{-\phantom{00}0.2}$ & $89^{+13}_{-27}$ & $\phantom{0}21^{+20}_{-11}$ & $0.07^{+0.03}_{-0.01}$ & $1.07^{+0.31}_{-0.37}$ & $1.32^{+0.07}_{-0.11}$ & $0.08^{+0.08}_{-0.06}$ & $137^{+27}_{-26}$ & $\phantom{00}6^{+\phantom{00}7}_{-\phantom{0}26}$ & $5$ & $1.0$ \\
81 & I &  & $5910^{+600}_{-700}$ & $46.1^{+\phantom{0}3.0}_{-\phantom{0}6.0}$ & $0.6^{+0.3}_{-0.1}$ & $\phantom{0}42\phantom{.0}^{+100\phantom{.0}}_{-\phantom{0}30\phantom{.0}}$ & $36^{+10}_{-10}$ & $\phantom{0}22^{+\phantom{0}3}_{-\phantom{0}3}$ & $0.15^{+0.04}_{-0.04}$ & $0.68^{+0.35}_{-0.13}$ & $1.13^{+0.04}_{-0.05}$ & $0.06^{+0.04}_{-0.03}$ & $111^{+\phantom{0}5}_{-\phantom{0}4}$ & $\phantom{00}7^{+\phantom{0}90}_{-\phantom{0}98}$ & $46$ & $1.9$ \\
83 & I &  & $4040^{+600}_{-400}$ & $44.6^{+\phantom{0}1.0}_{-\phantom{0}1.0}$ & $2.2^{+0.5}_{-0.3}$ & $\phantom{00}5.0^{+\phantom{00}1\phantom{.0}}_{-\phantom{00}1\phantom{.0}}$ & $84^{+17}_{-41}$ & $\phantom{0}49^{+10}_{-\phantom{0}7}$ & $0.14^{+0.04}_{-0.04}$ & $0.63^{+0.21}_{-0.10}$ & $1.33^{+0.05}_{-0.06}$ & $0.14^{+0.05}_{-0.07}$ & $\phantom{0}62^{+\phantom{0}9}_{-13}$ & $\phantom{0}51^{+\phantom{0}11}_{-\phantom{0}10}$ & $11$ & $1.7$ \\
85 & F &  & $4190^{+500}_{-300}$ & $42.6^{+\phantom{0}2.0}_{-\phantom{0}2.0}$ & $2.7^{+0.3}_{-0.4}$ & $\phantom{00}4.9^{+\phantom{0}20\phantom{.0}}_{-\phantom{00}2\phantom{.0}}$ & $40^{+32}_{-20}$ & $\phantom{0}16^{+\phantom{0}5}_{-\phantom{0}3}$ & $0.08^{+0.03}_{-0.02}$ & $0.86^{+0.39}_{-0.25}$ & $1.17^{+0.06}_{-0.06}$ & $0.12^{+0.06}_{-0.08}$ & $\phantom{0}35^{+20}_{-21}$ & $\phantom{0}76^{+\phantom{00}9}_{-119}$ & $8$ & $1.8$ \\
86 & F &  & $7730^{+200}_{-400}$ & $32.9^{+\phantom{0}0.7}_{-\phantom{0}0.7}$ & $2.7^{+0.2}_{-0.3}$ & $\phantom{00}0.6^{+\phantom{00}0.5}_{-\phantom{00}0.1}$ & $15^{+7}_{-3}$ & $\phantom{00}7^{+\phantom{0}2}_{-\phantom{0}1}$ & $0.16^{+0.03}_{-0.03}$ & $1.20^{+0.21}_{-0.35}$ & $1.38^{+0.08}_{-0.14}$ & $0.15^{+0.04}_{-0.07}$ & $127^{+31}_{-11}$ & $\phantom{0}23^{+168}_{-\phantom{00}5}$ & $18$ & $1.2$ \\
87 & I &  & $7920^{+60}_{-200}$ & $49.8^{+\phantom{0}0.2}_{-\phantom{0}0.4}$ & $1.8^{+0.1}_{-0.1}$ & $\phantom{00}0.2^{+\phantom{00}0.0}_{-\phantom{00}0.0}$ & $95^{+10}_{-19}$ & $\phantom{0}14^{+\phantom{0}3}_{-\phantom{0}2}$ & $0.19^{+0.01}_{-0.01}$ & $0.60^{+0.19}_{-0.07}$ & $1.30^{+0.05}_{-0.05}$ & $0.02^{+0.04}_{-0.02}$ & $169^{+\phantom{0}8}_{-11}$ & $\phantom{0}42^{+\phantom{0}16}_{-\phantom{0}56}$ & $48$ & $1.1$ \\
90 & I &  & $7710^{+200}_{-400}$ & $32.9^{+\phantom{0}3.0}_{-\phantom{0}4.0}$ & $2.1^{+0.4}_{-0.3}$ & $\phantom{00}0.6^{+\phantom{00}0.9}_{-\phantom{00}0.3}$ & $94^{+9}_{-17}$ & $\phantom{00}6^{+\phantom{0}2}_{-\phantom{0}1}$ & $0.19^{+0.01}_{-0.02}$ & $1.28^{+0.16}_{-0.35}$ & $1.11^{+0.12}_{-0.07}$ & $0.02^{+0.03}_{-0.02}$ & $\phantom{0}80^{+\phantom{0}2}_{-\phantom{0}1}$ & $\phantom{0}73^{+152}_{-\phantom{00}9}$ & $18$ & $0.9$ \\
94 & F &  & $7240^{+500}_{-700}$ & $31.5^{+\phantom{0}0.9}_{-\phantom{0}1.0}$ & $2.6^{+0.2}_{-0.4}$ & $\phantom{00}0.4^{+\phantom{00}0.1}_{-\phantom{00}0.1}$ & $79^{+19}_{-24}$ & $\phantom{0}21^{+12}_{-\phantom{0}6}$ & $0.08^{+0.04}_{-0.02}$ & $0.89^{+0.33}_{-0.29}$ & $1.12^{+0.04}_{-0.04}$ & $0.08^{+0.08}_{-0.05}$ & $\phantom{0}22^{+19}_{-13}$ & $\phantom{0}20^{+\phantom{0}38}_{-\phantom{0}64}$ & $53$ & $1.2$ \\
98 & II & K6    & $4020$ & $11.4^{+\phantom{0}0.4}_{-\phantom{0}0.4}$ & $2.8^{+0.1}_{-0.2}$ & $\phantom{0}11\phantom{.0}^{+\phantom{00}9\phantom{.0}}_{-\phantom{00}3\phantom{.0}}$ & $13^{+3}_{-1}$ & $\phantom{0}24^{+\phantom{0}3}_{-\phantom{0}2}$ & $0.10^{+0.01}_{-0.02}$ & $1.42^{+0.06}_{-0.10}$ & $1.02^{+0.02}_{-0.01}$ & $0.18^{+0.01}_{-0.03}$ & $127^{+\phantom{0}6}_{-\phantom{0}5}$ & $\phantom{0}99^{+179}_{-\phantom{0}44}$ & $47$ & $2.5$ \\
99 & I &  & $7440^{+400}_{-700}$ & $39.2^{+\phantom{0}2.0}_{-\phantom{0}1.0}$ & $0.9^{+0.2}_{-0.1}$ & $\phantom{00}0.4^{+\phantom{00}0.2}_{-\phantom{00}0.1}$ & $18^{+12}_{-5}$ & $\phantom{00}7^{+\phantom{0}4}_{-\phantom{0}2}$ & $0.13^{+0.04}_{-0.03}$ & $1.09^{+0.29}_{-0.38}$ & $1.36^{+0.08}_{-0.07}$ & $0.08^{+0.08}_{-0.05}$ & $\phantom{0}61^{+\phantom{0}9}_{-21}$ & $\phantom{00}6^{+\phantom{0}91}_{-\phantom{0}84}$ & $7$ & $1.0$ \\
100 & II & K5.5  & $4080$ & $11.6^{+\phantom{0}0.4}_{-\phantom{0}0.4}$ & $2.8^{+0.1}_{-0.2}$ & $\phantom{00}4.4^{+\phantom{00}4\phantom{.0}}_{-\phantom{00}1\phantom{.0}}$ & $25^{+22}_{-11}$ & $\phantom{0}17^{+\phantom{0}4}_{-\phantom{0}3}$ & $0.07^{+0.02}_{-0.01}$ & $1.08^{+0.29}_{-0.38}$ & $1.16^{+0.04}_{-0.03}$ & $0.12^{+0.06}_{-0.05}$ & $159^{+15}_{-19}$ & $\phantom{0}33^{+\phantom{0}65}_{-\phantom{0}88}$ & $4$ & $1.9$ \\
101 & II & K5    & $4140$ & $\phantom{0}5.3^{+\phantom{0}0.5}_{-\phantom{0}0.5}$ & $2.4^{+0.2}_{-0.2}$ & $\phantom{00}0.9^{+\phantom{00}0.2}_{-\phantom{00}0.2}$ & $26^{+26}_{-12}$ & $\phantom{0}43^{+19}_{-20}$ & $0.08^{+0.03}_{-0.02}$ & $1.23^{+0.19}_{-0.38}$ & $1.13^{+0.07}_{-0.04}$ & $0.16^{+0.03}_{-0.05}$ & $136^{+31}_{-29}$ & $\phantom{0}21^{+\phantom{0}37}_{-\phantom{0}72}$ & $6$ & $1.1$ \\
102 & I &  & $5210^{+400}_{-400}$ & $23.4^{+\phantom{0}0.8}_{-\phantom{0}0.8}$ & $0.6^{+0.1}_{-0.1}$ & $220\phantom{.0}^{+\phantom{0}80\phantom{.0}}_{-100\phantom{.0}}$ & $73^{+17}_{-15}$ & $\phantom{0}28^{+\phantom{0}3}_{-\phantom{0}3}$ & $0.18^{+0.01}_{-0.02}$ & $1.05^{+0.27}_{-0.30}$ & $1.47^{+0.02}_{-0.04}$ & $0.17^{+0.02}_{-0.05}$ & $121^{+\phantom{0}6}_{-\phantom{0}5}$ & $168^{+\phantom{0}10}_{-\phantom{00}8}$ & $17$ & $1.2$ \\
103 & II & K6    & $4020$ & $\phantom{0}3.6^{+\phantom{0}0.3}_{-\phantom{0}0.3}$ & $2.0^{+0.1}_{-0.1}$ & $\phantom{0}11\phantom{.0}^{+\phantom{00}3\phantom{.0}}_{-\phantom{00}2\phantom{.0}}$ & $16^{+9}_{-4}$ & $\phantom{0}74^{+\phantom{0}5}_{-30}$ & $0.05^{+0.01}_{-0.01}$ & $1.40^{+0.07}_{-0.13}$ & $1.09^{+0.02}_{-0.02}$ & $0.05^{+0.05}_{-0.03}$ & $165^{+11}_{-23}$ & $\phantom{0}68^{+\phantom{0}20}_{-\phantom{0}63}$ & $16$ & $2.7$ \\
104 & II & K5    & $4140$ & $\phantom{0}8.3^{+\phantom{0}0.4}_{-\phantom{0}0.4}$ & $2.4^{+0.1}_{-0.1}$ & $160\phantom{.0}^{+100\phantom{.0}}_{-100\phantom{.0}}$ & $82^{+18}_{-32}$ & $\phantom{0}22^{+\phantom{0}2}_{-\phantom{0}2}$ & $0.07^{+0.02}_{-0.01}$ & $0.83^{+0.39}_{-0.24}$ & $1.20^{+0.04}_{-0.04}$ & $0.13^{+0.05}_{-0.05}$ & $135^{+\phantom{0}9}_{-\phantom{0}7}$ & $133^{+\phantom{0}16}_{-\phantom{0}12}$ & $5$ & $3.0$ \\
107 & I &  & $3530^{+400}_{-300}$ & $28.6^{+\phantom{0}2.0}_{-\phantom{0}2.0}$ & $0.8^{+0.1}_{-0.1}$ & $\phantom{00}2.5^{+\phantom{00}0.8}_{-\phantom{00}0.5}$ & $42^{+16}_{-11}$ & $\phantom{0}24^{+11}_{-10}$ & $0.18^{+0.01}_{-0.03}$ & $1.00^{+0.31}_{-0.34}$ & $1.23^{+0.05}_{-0.05}$ & $0.00^{+0.01}_{-0.01}$ & $146^{+17}_{-15}$ & $\phantom{0}37^{+\phantom{0}52}_{-\phantom{0}79}$ & $23$ & $1.5$ \\
114 & II & M4    & $3160$ & $\phantom{0}6.6^{+\phantom{0}0.5}_{-\phantom{0}0.4}$ & $2.1^{+0.1}_{-0.1}$ & $\phantom{0}34\phantom{.0}^{+\phantom{0}10\phantom{.0}}_{-\phantom{00}7\phantom{.0}}$ & $28^{+24}_{-13}$ & $\phantom{0}58^{+\phantom{0}8}_{-\phantom{0}6}$ & $0.08^{+0.01}_{-0.01}$ & $0.97^{+0.35}_{-0.29}$ & $1.18^{+0.03}_{-0.02}$ & $0.01^{+0.02}_{-0.01}$ & $163^{+11}_{-13}$ & $\phantom{0}10^{+\phantom{0}98}_{-\phantom{0}47}$ & $2$ & $1.7$ \\
116 & II & K5    & $4140$ & $12.6^{+\phantom{0}1.0}_{-\phantom{0}1.0}$ & $2.4^{+0.2}_{-0.2}$ & $\phantom{00}4.3^{+\phantom{00}0.8}_{-\phantom{00}0.6}$ & $19^{+8}_{-6}$ & $\phantom{0}63^{+26}_{-16}$ & $0.07^{+0.02}_{-0.01}$ & $0.92^{+0.31}_{-0.30}$ & $1.11^{+0.03}_{-0.03}$ & $0.02^{+0.03}_{-0.01}$ & $113^{+12}_{-\phantom{0}6}$ & $\phantom{0}68^{+164}_{-176}$ & $7$ & $1.5$ \\
118 & II & M2    & $3490$ & $\phantom{0}5.2^{+\phantom{0}0.6}_{-\phantom{0}0.6}$ & $2.2^{+0.1}_{-0.1}$ & $\phantom{00}2.1^{+\phantom{00}0.4}_{-\phantom{00}0.4}$ & $91^{+12}_{-30}$ & $\phantom{0}56^{+13}_{-26}$ & $0.07^{+0.01}_{-0.01}$ & $1.21^{+0.21}_{-0.24}$ & $1.48^{+0.02}_{-0.03}$ & $0.17^{+0.02}_{-0.04}$ & $174^{+\phantom{0}5}_{-\phantom{0}9}$ & $\phantom{0}36^{+\phantom{0}41}_{-\phantom{0}83}$ & $12$ & $1.4$ \\
121 & II &  & $4610^{+800}_{-700}$ & $\phantom{0}5.5^{+\phantom{0}0.6}_{-\phantom{0}0.7}$ & $1.3^{+0.2}_{-0.2}$ & $\phantom{00}1.3^{+\phantom{00}0.3}_{-\phantom{00}0.3}$ & $64^{+28}_{-37}$ & $\phantom{0}37^{+23}_{-22}$ & $0.07^{+0.02}_{-0.01}$ & $1.29^{+0.16}_{-0.30}$ & $1.26^{+0.08}_{-0.06}$ & $0.15^{+0.03}_{-0.06}$ & $139^{+30}_{-29}$ & $\phantom{0}52^{+\phantom{0}21}_{-\phantom{0}87}$ & $5$ & $1.1$ \\
126 & F &  & $3130^{+200}_{-90}$ & $\phantom{0}0.7^{+\phantom{0}0.5}_{-\phantom{0}0.3}$ & $3.0^{+0.1}_{-0.1}$ & $\phantom{0}23\phantom{.0}^{+\phantom{00}8\phantom{.0}}_{-\phantom{00}6\phantom{.0}}$ & $12^{+2}_{-1}$ & $100^{+15}_{-25}$ & $0.11^{+0.04}_{-0.04}$ & $1.43^{+0.05}_{-0.09}$ & $1.45^{+0.03}_{-0.03}$ & $0.01^{+0.01}_{-0.01}$ & $\phantom{0}86^{+\phantom{0}7}_{-\phantom{0}1}$ & $\phantom{0}48^{+\phantom{00}3}_{-\phantom{00}3}$ & $62$ & $1.4$ \\
130 & I &  & $6680^{+700}_{-800}$ & $30.7^{+\phantom{0}1.0}_{-\phantom{0}1.0}$ & $2.2^{+0.5}_{-0.5}$ & $\phantom{0}38\phantom{.0}^{+\phantom{00}9\phantom{.0}}_{-\phantom{00}8\phantom{.0}}$ & $41^{+23}_{-17}$ & $\phantom{0}59^{+\phantom{0}7}_{-\phantom{0}6}$ & $0.16^{+0.03}_{-0.04}$ & $1.39^{+0.08}_{-0.14}$ & $1.19^{+0.03}_{-0.03}$ & $0.01^{+0.02}_{-0.01}$ & $\phantom{0}48^{+\phantom{0}4}_{-\phantom{0}4}$ & $142^{+\phantom{00}6}_{-\phantom{00}6}$ & $58$ & $1.8$ \\
\hline
\end{tabular}
\renewcommand{\arraystretch}{1.0}
\end{table}
\end{landscape}

\begin{landscape}
\begin{table}
\centering
\renewcommand{\arraystretch}{1.4}
\begin{tabular}{lllllllllllllllll}
\hline
ID & Spec. & Spec. & $T_\mathrm{eff}$ & $A_V$ & $R_*$ & $M_\mathrm{dust}$ & $r_\mathrm{in}$ & $r_\mathrm{out}$ & $h/r$ & $\gamma$ & $\beta$ & $\xi$ & $i$ & $\mathrm{PA}$ & $\chi^2_\mathrm{SED}$ & $\chi^2_\mathrm{img}$ \\
 & class & type & (K) & (mag) & ($R_\odot$) & ($M_\oplus$) & ($R_\odot$) & (au) & & & & & ($^\circ$) & ($^\circ$) & & \\
\hline
131 & I &  & $3640^{+500}_{-200}$ & $10.0^{+\phantom{0}0.5}_{-\phantom{0}0.4}$ & $2.9^{+0.1}_{-0.1}$ & $\phantom{00}2.6^{+\phantom{00}0.6}_{-\phantom{00}0.6}$ & $40^{+10}_{-10}$ & $\phantom{0}41^{+\phantom{0}9}_{-\phantom{0}7}$ & $0.14^{+0.02}_{-0.06}$ & $1.42^{+0.06}_{-0.07}$ & $1.45^{+0.03}_{-0.03}$ & $0.01^{+0.01}_{-0.01}$ & $\phantom{0}84^{+\phantom{0}1}_{-\phantom{0}1}$ & $\phantom{0}94^{+\phantom{0}11}_{-\phantom{0}11}$ & $46$ & $1.8$ \\
132 & I &  & $5740^{+400}_{-600}$ & $20.6^{+\phantom{0}0.8}_{-\phantom{0}0.8}$ & $1.2^{+0.2}_{-0.1}$ & $\phantom{00}0.6^{+\phantom{00}0.2}_{-\phantom{00}0.1}$ & $86^{+14}_{-17}$ & $\phantom{00}6^{+\phantom{0}2}_{-\phantom{0}1}$ & $0.14^{+0.03}_{-0.04}$ & $1.32^{+0.13}_{-0.31}$ & $1.21^{+0.10}_{-0.08}$ & $0.00^{+0.01}_{-0.01}$ & $154^{+19}_{-26}$ & $\phantom{0}51^{+\phantom{00}9}_{-\phantom{0}89}$ & $19$ & $1.2$ \\
133 & II &  & $3860^{+300}_{-300}$ & $14.7^{+\phantom{0}0.6}_{-\phantom{0}0.7}$ & $1.9^{+0.3}_{-0.2}$ & $\phantom{00}0.9^{+\phantom{00}0.2}_{-\phantom{00}0.2}$ & $85^{+15}_{-29}$ & $\phantom{0}15^{+\phantom{0}9}_{-\phantom{0}5}$ & $0.06^{+0.02}_{-0.01}$ & $1.05^{+0.31}_{-0.35}$ & $1.32^{+0.07}_{-0.07}$ & $0.08^{+0.07}_{-0.05}$ & $155^{+18}_{-23}$ & $\phantom{0}36^{+\phantom{0}59}_{-\phantom{0}80}$ & $4$ & $1.1$ \\
139 & F &  & $6440^{+700}_{-2000}$ & $12.5^{+\phantom{0}1.0}_{-\phantom{0}0.9}$ & $0.9^{+0.8}_{-0.2}$ & $\phantom{00}9.7^{+\phantom{0}30\phantom{.0}}_{-\phantom{00}8\phantom{.0}}$ & $21^{+12}_{-6}$ & $\phantom{00}7^{+\phantom{0}2}_{-\phantom{0}1}$ & $0.16^{+0.03}_{-0.03}$ & $0.87^{+0.37}_{-0.26}$ & $1.20^{+0.04}_{-0.04}$ & $0.10^{+0.06}_{-0.05}$ & $\phantom{0}57^{+\phantom{0}6}_{-\phantom{0}7}$ & $\phantom{0}57^{+156}_{-\phantom{0}86}$ & $10$ & $1.5$ \\
144 & F &  & $3190^{+200}_{-100}$ & $\phantom{0}9.9^{+\phantom{0}1.0}_{-\phantom{0}1.0}$ & $1.9^{+0.3}_{-0.3}$ & $\phantom{00}2.0^{+\phantom{00}0.5}_{-\phantom{00}0.4}$ & $45^{+23}_{-21}$ & $\phantom{0}65^{+27}_{-18}$ & $0.14^{+0.03}_{-0.03}$ & $1.28^{+0.14}_{-0.25}$ & $1.43^{+0.05}_{-0.06}$ & $0.01^{+0.01}_{-0.01}$ & $\phantom{0}89^{+\phantom{0}3}_{-\phantom{0}3}$ & $107^{+\phantom{0}37}_{-\phantom{0}38}$ & $14$ & $0.9$ \\
\hline
\end{tabular}
\renewcommand{\arraystretch}{1.0}
\end{table}
\end{landscape}

\bsp	
\label{lastpage}
\end{document}